\documentclass[twocolumn,astrosymb]{aastex631}
\usepackage{rotating}
\usepackage{booktabs}
\usepackage{textcomp, gensymb}
\usepackage{tabularx}
\usepackage{amsmath}
\shorttitle{Spotted WTTS in Taurus-Auriga}
\shortauthors{P\'erez Paolino et al.}
\graphicspath{{./}{figures/}}

\begin{document}

\title{The Effect of Starspots on Spectroscopic Age and Mass Estimates of Non-Accreting T~Tauri Stars in the Taurus-Auriga Star Forming Region}

\correspondingauthor{Jeff Bary}
\email{jbary@colgate.edu}

\author[0000-0002-4128-7867]{Facundo P\'erez Paolino}
\affiliation{Colgate University, 13 Oak Drive, Hamilton, NY 13346, USA}
\affiliation{Department of Astronomy, California Institute of Technology, Pasadena, CA 91125, USA}
\affiliation{Visiting Astronomer at the NASA Infrared Telescope Facility}
\altaffiliation{NASA's Infrared Telescope Facility is operated by the University of Hawaii under contract 80HQTR19D0030 with the National Aeronautics and Space Administration.}

\author[0000-0001-8642-5867]{Jeffrey S. Bary}
\affiliation{Colgate University, 13 Oak Drive, Hamilton, NY 13346, USA}
\affiliation{Visiting Astronomer at the NASA Infrared Telescope Facility}

\author{Lynne A. Hillenbrand}
\affiliation{Department of Astronomy, California Institute of Technology, Pasadena, CA 91125, USA}

\author[0009-0000-9670-2194]{Madison Markham}
\affiliation{Colgate University, 13 Oak Drive, Hamilton, NY 13346, USA}
\affiliation{Visiting Astronomer at the NASA Infrared Telescope Facility}




\begin{abstract}
Accurate age and mass determinations for young pre-main sequence stars are made challenging by the presence of large-scale starspots. We present results from a near-infrared spectroscopic survey of ten T-Tauri Stars in Taurus-Auriga that characterize spot filling factors and temperatures, the resulting effects on temperature and luminosity determinations, and the consequences for inferred stellar masses and ages. We constructed composite models of spotted stars by combining BTSettl-CIFIST synthetic spectra of atmospheres to represent the spots and the photosphere along with continuum emission from a warm inner disk. Using a Markov-Chain Monte-Carlo algorithm, we find the best-fit spot and photospheric temperatures, spot filling factors, as well as disk filling factors. This methodology allowed us to reproduce the 0.75-2.40 micron stellar spectra and molecular feature strengths for all of our targets, disentangling the complicated multi-component emission. For a subset of stars with multi-epoch observations spanning an entire stellar rotation, we correlate the spectral variability and changes in the filling factors with rotational periods observed in K2 and AAVSO photometry. Combining spot-corrected effective temperatures and Gaia distances, we calculate luminosities and use the Stellar Parameters of Tracks with Starspots (SPOTS) models to infer spot-corrected masses and ages for our sample of stars. Our method of accounting for spots results in an average increase of 60\% in mass and a doubling in age with respect to traditional methods using optical spectra that do not account for the effect of spots.
\end{abstract}



\section{Introduction} \label{sec:intro}

{Pre-Main-Sequence (PMS) stars are characterized by strong magnetic fields and rapid rotation rates, leading to strong magnetic flares and heightened chromospheric and coronal activity \citep{Berdyugina2005, Toriumi2019}. These strong magnetic fields can also guide continued accretion flows onto the stellar surfaces resulting in hot-spots and thus excess continuum emission \citep{Hartmann2016}. Further complicating the study of these systems is the possibility that convection near the surface of the star may be suppressed by strong magnetic fields, resulting in the formation of starspots with temperatures cooler than the photosphere covering significant fractions of the stellar surface \citep{Rydgren&Vrba1983,Bary2014,Gully2017,myself}. Such large-scale starspot activity makes the stars appear redder and complicates the measurement of stellar parameters such as radius, effective temperature, and luminosity \citep{Berdyugina2005, Guo2018}. Modeling the effects of starspots and disentangling this multi-component emission is therefore essential for understanding young PMS stars}. 

{Several studies \citep[e.g.,][]{Debes2013,Bary2014,Gully2017,myself} have shown that the near-infrared (NIR) spectra of some young T~Tauri Stars (TTS) are best reproduced not by any single-temperature spectral type, but by a weighted combination of two different temperature components, a hotter one representing the unspotted photosphere, and a cooler one representing the starspots}. While these studies have been successful in reconstructing narrow regions of the stellar spectra, it still remains a challenge to simultaneously fit broader spectral windows. \citet{Gully2017} used composite two-temperature models to fit high-resolution NIR IGRINS spectra of LkCa~4 and found evidence for spot complexes covering up to 80\% of the stellar surface. Using conventional evolutionary models that do not take into account the presence of spots, they find the mass of LkCa~4 to differ by as much as two to four times when compared to studies using single-temperature models \citep[e.g.][]{Donati2014}. {This represents a decrease in the stellar mass from $0.79\pm{0.05}\ M_{\odot}$ to $0.15-0.30\ M_{\odot}$.} \citet{Gangi} performed a larger survey of young stars in Taurus-Auriga and found systematic effects on age and mass determinations when using single-temperature models as opposed to two-temperature models for studying spotted stars. {Cumulatively, these studies highlight the systematic effects that failing to account for starspots can have on determinations of the effective temperatures and luminosities of young stars as well as on mass and age inferences}. In addition, starspots appear to be a reasonable explanation for the spectral type mismatches seen in young stars \citep{Gullbring1998, vacca2011}. 

Mass inference for PMS stars typically relies on a direct comparison between observationally-determined temperatures and luminosities to those predicted in theoretical evolutionary models \citep[e.g.,][]{DAntona1994,Baraffe2015,Soderblom2014}. {While this technique has shown great success in older stellar populations, there are still large mismatches present in the inferred ages and masses of young stars \citep{Hartmann2001}. Similar issues are seen between spectral types inferred from infrared and optical observations \citep[e.g., ][]{vacca2011, Bary2014, kastner2015}, and in color anomalies \citep{Gullbring1998, Pecaut2016} for pre-main-sequence stars.} \citet{Flores2022} compares dynamical mass measurements to masses inferred using infrared and optical spectroscopy for 24 young stars in Taurus-Auriga and Ophiuchus, finding that mass inferences using infrared spectra provide the most reliable results, overestimating stellar masses by only 4\% over the entire mass range. Inferences based on optical spectra on the other hand provide far less reliable results, overestimating stellar masses by 36\% over the entire mass range, with the difference becoming larger for stars under 0.5 solar masses where optical temperatures over-predict stellar masses by 94\%. 

Recent work by \citet{myself} used multi-epoch NIR spectra of LkCa~4 to find a correlation between variations in the starspot filling factors and photometric variability spanning more than one rotation of the star. These results provide strong evidence for large-scale starspot coverage being responsible for the observed photometric variability seen Weak-Lined T Tauri Stars (WTTS) and Classical T Tauri Stars (CTTS). Furthermore, their models suggest spots covering {on the order of} $\approx85\%$ of the stellar surface in line with those predicted from photometry alone by \citet{Grankin2008}. In this scenario, the presence of spots would substantially alter the effective temperatures measured for these stars as well as their spectral type, so the authors use new updated evolutionary models from \citet{Somers2020} to study the inherent uncertainty in mass determinations arising from the use of single-epoch spectral observations of spotted stars and find uncertainties in the order of 20\% for LkCa 4. While the adoption of two-temperature fits serves to {mitigate} the uncertainties generated by spots, these results {indicate} the need for future studies of larger stellar populations using infrared multi-epoch spectroscopy to {more} accurately measure the ages and masses of young spotted stars. {Such studies will require models that account for emission from accretion hot spots ($T\approx$~8000~K) to model actively accreting CTTS \citep[e.g., ][]{Fischer2011}. In addition, near-infrared excess emission from warm circumstellar dust ($T\approx$~1500~K) in the inner regions of both CTTS and WTTS disks will need to be included in the models.}

In this work, we perform a similar analysis to \citet{myself} of NIR spectra for a sample of ten non-accreting WTTS in the Taurus-Auriga star forming region. {Using BTSettl-CIFIST \citep{allard2014} theoretical atmospheres, we build spectral models that simultaneously account for the stellar photosphere, starspot, and a blackbody component to model the disk emission.} Our composite spectra over the 0.75-2.40~\micron\ window reproduce the observed spectra and constrain spot temperatures and filling factors. 

This article is structured as follows. In Section~\ref{sec:obs} we outline our observations, reduction process, and approximate flux-calibration. In section~\ref{sec:spotmod}, we compare our reduced spectra to main sequence standards and look for evidence of surface starspot coverage. Following this, in Section~\ref{sec:avteff} we derive a self consistent set of what we will refer to as Red-NIR temperatures and extinctions for stars in our sample. In Section~\ref{sec:models}, we introduce our models of spotted stars and fit every epoch of available NIR SpeX spectra to constrain spot filling factors and temperatures as well as photospheric temperatures. We use use the available K2 photometry for a subset of stars in our sample to correlate photometric variability with spectroscopic variability. In Section~\ref{sec:HR} we use our constraints on the spot properties along with Gaia EDR3 distances to calculate spot-based stellar luminosities and effective temperatures for the stars in our sample. Placing the stars in an HR~Diagram alongside spotted and un-spotted evolutionary models, we infer ages and masses from optical, Red-NIR, and spot corrected effective temperatures and luminosities. In section~\ref{sec:discussion} we discuss our results in the broader context of starspots on young stars and examine the limitations of our method. Finally, in Section~\ref{sec:conclusion} we summarize our results. Incorporating accretion onto our models of spotted stars is beyond the scope of this paper, but reserved for an upcoming study of spotted accreting CTTS in Taurus-Auriga.

\section{Observations} \label{sec:obs}
Our data {comes} from two different data sets. Spectra were obtained using SpeX, a medium-resolution cross-dispersed near-IR spectrograph at NASA's Infrared Telescope Facility (IRTF) over two observing runs in 2019 and 2022. {For both of these runs, we used the short-wavelength cross-dispersed mode \citep[SXD;][]{rayner2003} with the 0.3\arcsec $\times$ 15\arcsec~slit (R$\sim$2000)}. A single spectrum of V819~Tau was acquired in 2006 by William Fischer and Suzan Edwards, and published in \citet{Fischer2011}. This spectrum, which displayed clear signatures of starspot contamination, was obtained through private communication.

The 2019 and 2022 observing runs benefited from the upgraded detector in SpeX following its August 2014 update. The upgraded Teledyne 2048x2048 Hawaii-2RG array with its 0.10\arcsec pixel size provides continuous spectral coverage between 0.70-2.55~\micron\, where the old Raytheon Aladdin 3 1024x1024 InSb array with its 0.5\arcsec\ pixel size provided continuous spectral coverage across 0.80-2.40~\micron. This resulted in a 0.25~\micron\ increase in the spectral coverage between 2006 and the most recent observing runs. {During the 2006 observing run exclusively single epoch spectra were acquired, while during the 2019 and 2022 observing runs multi-epoch spectral coverage over five to eight nights for most of the targets was obtained multi-epoch spectral coverage over five to eight nights for most of the targets, with a handful of single epoch observations.} Both the 2019 and 2022 observing runs benefited from excellent weather conditions leading to a typical 0.6-1.0\arcsec seeing.

The data were collected using a standard ABBA nod sequence. The subtraction of AB and BA image pairs allows for the efficient removal of terrestrial OH emission lines, background, and dark current. A0V telluric standards were observed with an airmass difference of $\Delta \sec z~\leq0.1$ to the target and were used to remove telluric absorption features as well as for an approximate flux calibration. Flat-field corrections, wavelength calibrations, spectral extraction, coaddition, telluric corrections, and merging of the spectral orders were performed using the IDL-based reduction package Spextool \citep{cushing2004}. Typical signal to noise ratios are 100--300 for all observations.

\begin{deluxetable*}{c|ccccc|ccc|ccc} 
\tablecaption{Stellar Sample, Multiplicity, Distances, Disks, Rotational Periods, Single-Temperature Fit Parameters and Corresponding Stellar Parameters.}\label{tab:obsdata}
\tablehead{
\colhead{Star} & \colhead{$N_{\mathrm{obs}}$} & \colhead{Multiple$^{a}$} & \colhead{Distance$^{b}$ [pc]} & \colhead{$P_{\mathrm{rot}}$ [days]} & \colhead{Ref} &\colhead{$T_{\mathrm{red}}$ [K]} & \colhead{$\log\ (L_{\mathrm{red}}/L_{\odot})$} & \colhead{$A_{V}$ [mags]} & \colhead{$T_{\mathrm{opt}}$ [K]} & \colhead{$\log\ (L_{\mathrm{opt}}/L_{\odot})$} & \colhead{Ref}\\
\colhead{} & \colhead{} & \colhead{} & \colhead{} & \colhead{} &\colhead{} & \colhead{} & \colhead{} & \colhead{} & \colhead{} & \colhead{} & \colhead{}}
\setlength{\tabcolsep}{4pt}
\small
\decimals
\startdata
DI Tau & 4 & Binary & $137.6\pm{0.9}$ & \nodata & \nodata & $3783\pm{71}$ & $-0.14\pm{0.04}$ & $0.83\pm{0.28}$ & 3560 & -0.15 & 6\\ 
IW Tau & 14 & Binary & $142.0\pm{0.5}$ & $5.50$, $7.04$ & 2 &  $3713\pm{49}$ & $-0.11\pm{0.07}$ & $0.63\pm{0.14}$ & 4100 & -0.12 & 6\\ 
JH 108 & 10 &  Single & $162.4\pm{0.5}$ & $6.53$ & 3 &  $3659\pm{44}$ & $-0.11\pm{0.07}$ & $1.66\pm{0.18}$ & 3660 & -0.35 & 6\\
LkCa 4 & 12 &  Single & $129.8\pm{0.3}$ & $3.374$ & 4 &  $3650\pm{73}$ & $-0.35\pm{0.08}$ & $0.70\pm{0.21}$ & 4060 & 0.02 & 6\\
LkCa 7 & 1 &  Single & $126.4\pm{1.6}$ & $5.66$ & 4 &  $3811\pm{70}$ & $-0.21\pm{0.10}$ & $0.15\pm{0.08}$ & 4100 & -0.23 & 6\\
LkCa 14 & 4 &  Single & $128.2\pm{0.5}$ & \nodata & \nodata &  $4105\pm{60}$ & $-0.22\pm{0.06}$ & $0.08\pm{0.05}$ & 3850 & -0.20 & 6\\
V819 Tau & 1 &  Single & $129.3\pm{0.4}$ & $5.53$ & 4 &  $3797\pm{96}$ & $-0.19\pm{0.06}$ & $1.29\pm{0.20}$ & 4060 & 0.02 & 5\\
V826 Tau & 13 & Binary & $143.7\pm{0.4}$ & $3.88$ & 1 &  $3897\pm{51}$ & $0.02\pm{0.06}$ & $0.29\pm{0.17}$ & 4100 & -0.01 & 6\\
V830 Tau & 10 &  Single & $130.4\pm{0.3}$ & $2.74$ & 2 &  $3874\pm{57}$ & $-0.26\pm{0.05}$ & $0.41\pm{0.20}$ & 4100 & -0.27 & 6\\
V1075 Tau & 8 &  Single & $143.5\pm{0.4}$ & $2.43$ & 2 &  $4021\pm{82}$ & $-0.19\pm{0.04}$ & $0.17\pm{0.08}$ & 4100 & 0.21 & 6\\ 
\enddata
\vspace{0.1cm}
(a)~Multiplicity from \citet{Gangi}. (b)~Distances from \textit{Gaia} EDR 3 \citep{Gaia}. 
(1)~\citet{Cody2022}. (2)~\citet{Rebull2020}. (3)~Computed via Lomb-Scargle periodogram in the NASA Exoplanet Archive time series tool using K2 photometry \citep{Akeson2013}. (4)~\citet{Grankin2008}. (5)~\citet{andrews2013}. (6)~\citet{Dent2013}.
\end{deluxetable*}

\section{Spectroscopic Evidence for Spots in WTTS}\label{sec:spotmod}
\begin{figure*}[!ht]
\centering
\includegraphics[width=1\linewidth]{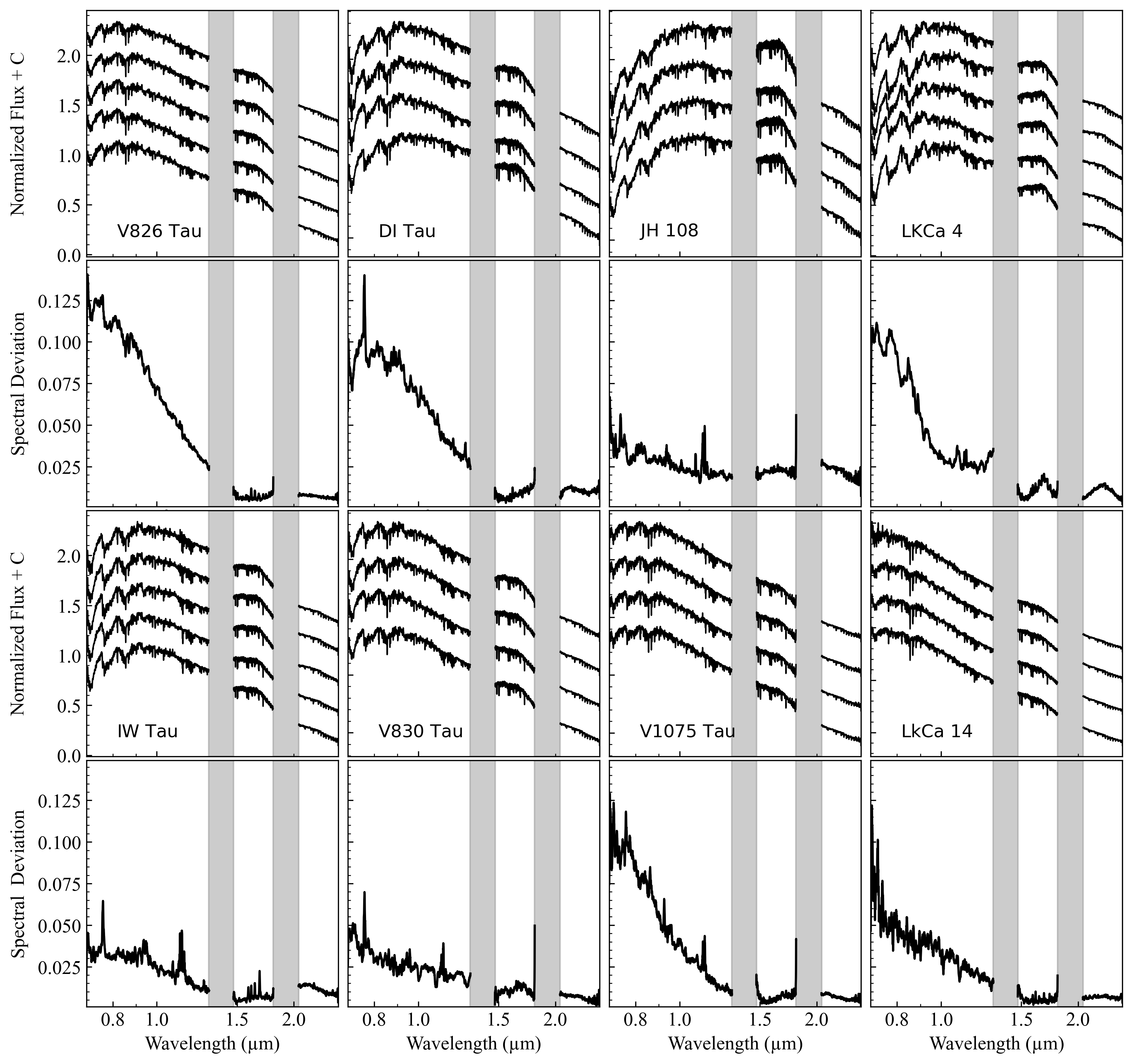}
\caption{Plotted are normalized spectra from every night of observations for V826~Tau, DI~Tau, JH~108, and LkCa~4 (First row), IW~Tau, V830~Tau, V1075~Tau, and LkCa~14 (Third row) along with their corresponding deviation spectrum (Second and fourth rows). All spectra are plotted in chronological order, with the first night spectra at the top. Areas shaded in grey have been cut due to poor atmospheric transmission. For display purposes, the spectra have been smoothed using a three-pixel boxcar.}
\label{fig:spectraobservations}
\end{figure*}

Inspection of our multi-epoch spectroscopic data reveals clear and pronounced night-to-night variability in both the Spectral Energy Distribution (SED) of the stars and in the relative strengths of several molecular features in the I and Y band, most notably Titanium Oxide (TiO) and Iron Hydride (FeH). Figure~\ref{fig:spectraobservations} shows normalized spectra of four stars spanning four and five observations, where we only show the first spectra of each night. Throughout, we use IYJHK to refer to the standard atmospheric windows \citep{Mann2015} that are encompassed by the Spex spectral coverage.

To study the wavelength dependence of the spectral variability, we have computed deviation spectra for all stars for which we possess multi-epoch data. The deviation spectra were computed from the standard deviation across the flux values with respect to the median from every epoch at each wavelength element over the entire 0.70-2.55 \micron\ or 0.80-2.40 \micron\ spectral window depending on the available spectral coverage. The final deviation spectra can be found directly beneath the corresponding stellar spectra in Figure~\ref{fig:spectraobservations}. Across all wavelengths, we see spectral variability between one and fifteen percent, but certain molecular and atomic features in the spectra show higher levels of variability than the surrounding continua. In particular, the TiO bands at $\approx0.75$~\micron\ and $\approx0.85$~\micron, the FeH feature at $\approx0.99$~\micron\, and the H-band ``hump" between $\approx1.60$-1.75~\micron\ display enhanced variability with respect to the stellar continuum. 

Several studies have used TiO and FeH as a proxy for starspots on the surfaces of young stars \citep{Bary2014,Gully2017,myself}, and these results support a starspot-based explanation of the observed spectral variability, in which the features appear stronger in observations taken when a larger fraction of the visible hemisphere of the star was  covered with spots. Not only are starspots colder, but owing to their lower temperature, also possess different molecular abundances and radiative properties with respect to the surrounding photosphere. 

These results, in combination with those of \citet{Fischer2011} suggest that the region of the spectra of spotted stars most suitable for decomposing the complicated emission from the combination of photosphere, spots, and possibly cold disk and/or hot accretion components lies between 0.70-1.0~\micron. This region appears to be the least contaminated by veiling from either the disk or the accretion activity and contains the most significant spectroscopic spot signatures.

\begin{figure*}[!ht]
\centering
\includegraphics[width=1\linewidth]{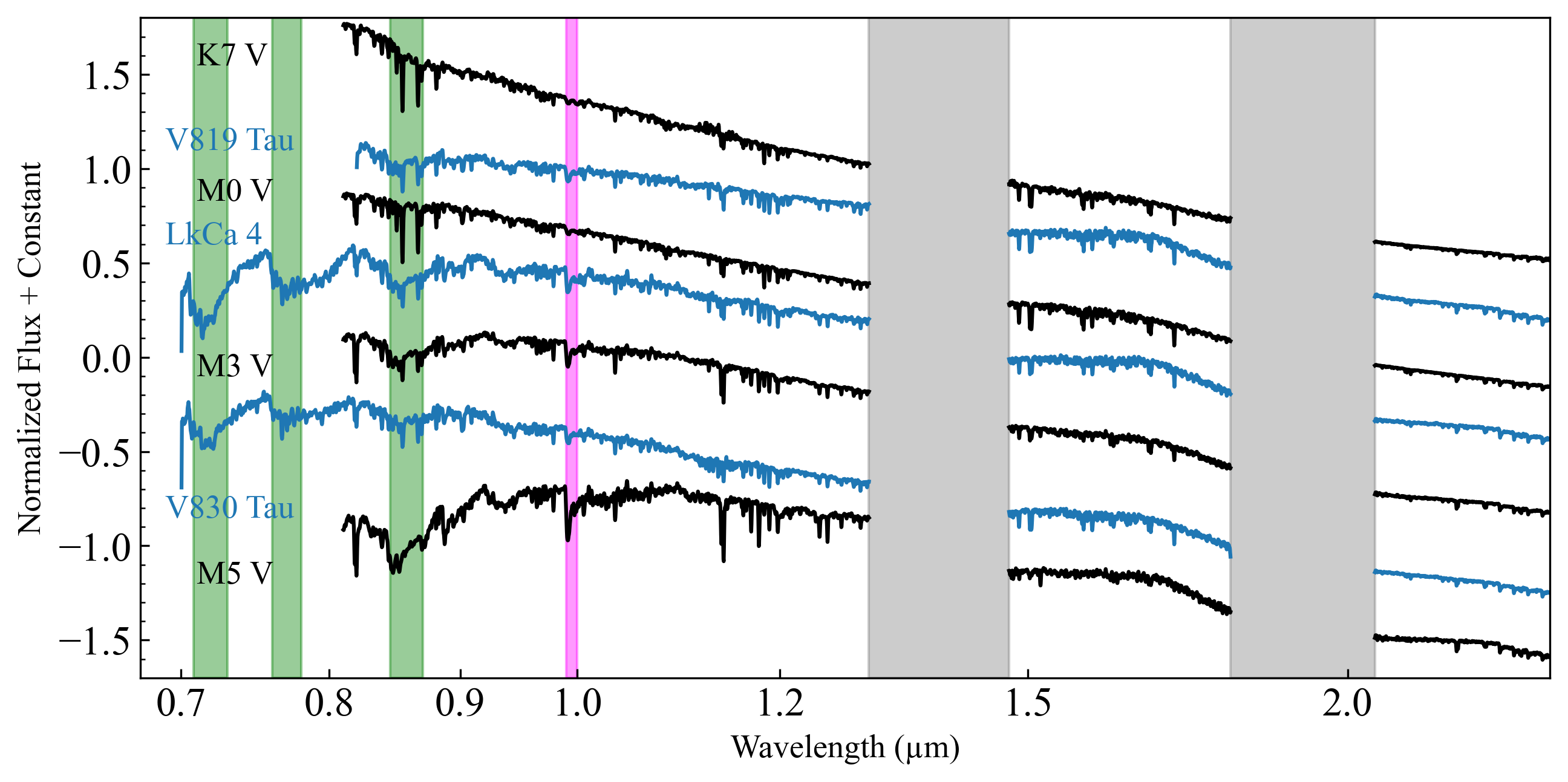}
\caption{Comparison between dereddened normalized V819 Tau, LkCa~4, and V830 Tau spectra (blue) and Spex Spectral Library standard stars of spectral type K7V, M0V, M3V, and M5V (black). Areas shaded in grey have been cut due to poor atmospheric transmission. The spectra have been smoothed in the plot in a three pixel window and offset by a constant for display purposes. Some spectra are displayed down to 0.82~\micron\ since that is the cutoff for SpeX prior to its upgrade in 2014. The location of the strongest TiO band has been highlighted in green, while the FeH band is indicated in magenta.}
\label{fig:lkca4comparisonsingle}
\end{figure*}

While V819~Tau, LkCa~4, and V830~Tau possess literature spectral types of K7V, K7V, and M0V, respectively \citep{Feigelson1981, Donati2014, Strassmeier2009}, as derived from optical spectra, their near-infrared spectra show several features that are characteristic of much cooler M-dwarfs. This makes their classification troublesome with single temperature fits. Figure~\ref{fig:lkca4comparisonsingle} shows a comparison between dereddened spectra of (see section \ref{sec:avteff}) V819~Tau, LkCa~4, and V830~Tau  and the SpeX spectral standards, K7V (HD~201092 - $T_{eff}=4100\ K$), M0V (HD~19305 - $T_{eff}=3850\ K$), M3V (Gl~388 - $T_{eff}=3430\ K$), and M5V (Gl~51 - $T_{eff}=3060\ K$). For wavelengths $\leq0.9\ \micron$, the K7V and M0V standards possess much shallower TiO and FeH molecular features than any of the WTTS. Contrary to this, the calcium triplet near $\approx0.85\ \micron$ and the sodium doublet near $\approx0.82\ \micron$ are much stronger in the K7V and M0V standards. Across the entire spectrum, the SEDs of the WTTS are not closely matched by either the M0V or the K7V standards, nor by the much cooler M5V spectra, but most closely matched by the M3V. The differences continue at longer wavelengths, where at H-band the M5V and not the M3V most closely reproduces the broad hump-like feature. 

These features are not unique to the stars shown, but are present in all stars in our sample to varying degrees (see Figure~\ref{fig:spectraobservations}), indicating the contribution of cooler photospheric components to fully explain the SED and molecular features in our sample of WTTS. Therefore, the inability of single-temperature template spectra to simultaneously reproduce both the shape of the continuum and the strength of the molecular and atomic features also strongly suggests that the stellar spectra are not produced by single temperature photospheres, but from ones with at least two components with significantly different temperatures. The differences in the strengths of the atomic and molecular features also rule out the possibility of the discrepancies being the result of a strong veiling continuum from an unseen disk or undetected accretion in these systems, as this would reduce the strength of the features for both the atomic lines and the molecular bands, but not their relative strengths.

\section{Measuring Extinction and Temperature} \label{sec:avteff}
The results of \citet{Fischer2011} suggest that the excess veiling continuum originating from accretion and disk emission in TTS spectra is minimized in the region between 0.75-1.00~\micron. For wavelengths shorter than 0.70~\micron, emission from accreting gas with temperatures between 8000-10000~K can dominate the stellar continuum for high mass accretion rates. For wavelengths longer than 1.00~\micron, the emission from warm dust ($T=1400-2000~K$) in the inner disk becomes significant, with its relative strength depending on the location and geometry. 

In light of this minimal contribution from any circumstellar source of excess emission, we use the 0.75-1.00~\micron\ region of the NIR SpeX spectra to simultaneously determine $A_{V}$ and $T_{eff}$. We avoid the first 0.05~\micron\ because the signal-to-noise drops significantly at the edges of the detector. To measure $A_{V}$ and $T_{eff}$, we use an emcee-based \citep{Foreman-Mackey2013} Markov-Chain Monte-Carlo algorithm to fit the stellar spectrum to a reddened BTSett-CIFIST model atmosphere at $\log g$~=~3.5. The grid of theoretical spectra was first re-sampled using the simple, flux-conserving re-sampler SpectRes~\citep{2017arXiv170505165C}, degraded to SpeX resolution using Coronagraph \citep{Coronagraph1, Coronagraph2}, and then interpolated to provide continuous temperature coverage between 1200~K and 7000~K. The spectra were reddened following the extinction law of \citet{Fitzpatrick1999} with an $R_{V}=5.5$ {following \citet{Gangi} to allow for a direct comparison between both methods of determining $A_V$.} The resulting posterior parameter distribution was then used to find the highest-likelihood solution, as well as to determine a $1\sigma$\ uncertainty. We repeated this process across every epoch of spectral data, finding that $T_{eff}$ and $A_{V}$ showed small ($\leq20~K$ and $\leq$ 0.1~mags) variations between epochs, so we calculated an average and standard deviation across all epochs and adopt them as our final extinctions and temperatures. 

\begin{figure}[!ht]
\centering
\includegraphics[width=1\linewidth]{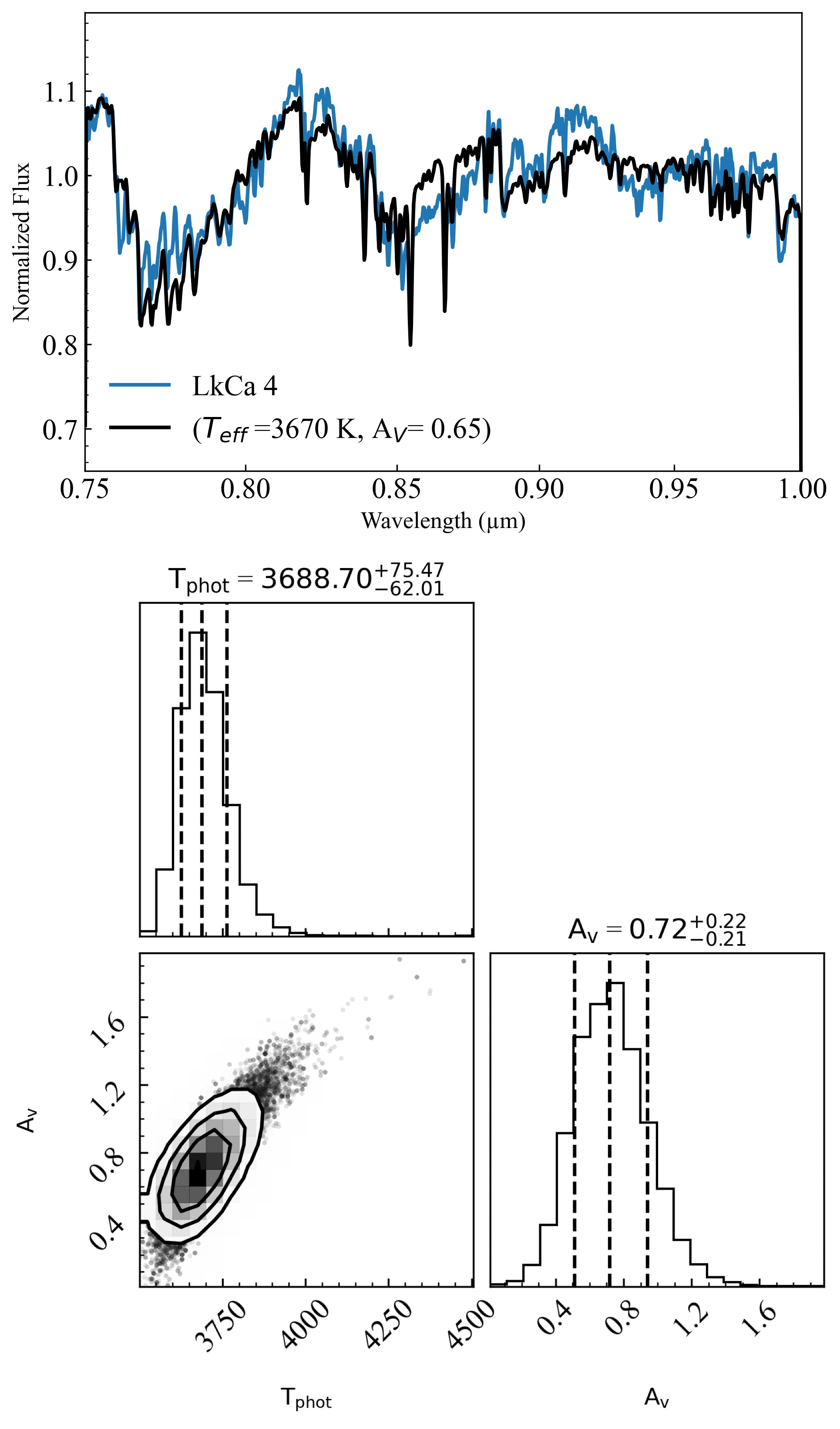}
\caption{Top: Best-fit reddened single-temperature BTSettl-CIFIST atmosphere (black) compared to the corresponding LkCa~4 spectrum (MJD~58489.214,~blue). The LkCa~4 and model spectra have been smoothed for display purposes. Bottom: Corner plot showing the posterior parameter distribution for the $A_{V}$ fit. The minor difference in reported parameters is due to the top figure reporting maximum-likelihood values calculated from the posteriors, while on the bottom shown median values for the posterior distributions.}
\label{fig:avfit}
\end{figure}

Results are shown in Table~\ref{tab:obsdata}, while an example of an $A_{v}$ fit to a single LkCa~4 spectrum and its corresponding posterior parameter distribution, median, and $1\sigma$ uncertainties are shown in Figure~\ref{fig:avfit}. Typical extinctions for stars in our sample are between 0.3-0.8 mags, with values in closer agreement with extinctions determined in the optical \citep[e.g., ][]{herczeg2014}, but systematically lower than those determined further in the NIR \citep[e.g., ][]{Kenyon1995, Gullbring1998}. It should be noted that both of these studies used a lower $R_{V}=3.1$. Given the presence of strong TiO absorption bands in this portion of the spectrum and their likely spot origin, we expect temperatures determined in this spectral region to be shifted towards colder values with respect to the photospheric temperature. For regions at shorter wavelengths, we expect the effective temperature to asymptotically approach that of the stellar photosphere. These estimates then represent a fairly reliable indicator of the effective temperature of a spotted star, and with $A_{V}$ constraints less biased than those derived from similar fitting at shorter and longer wavelengths. We refer to these temperatures as Red-NIR temperatures $T_{red}$ due to the spectral window used in their inference and because of the strong TiO absorption bands present. As a single temperature fit, these values represent an effective temperature that is an average of the photospheric and spot temperatures.

{In cases where the posteriors for $T_{eff}$ and $A_{V}$ possessed a long tail extending out to unreasonably high values, we truncate the parameter space and focus on values of $A_{V}<2$ and $T_{eff}\leq5000$~K. We justify these restrictions based on the lack of high-excitation atomic and ionization features, which would be present in the spectra of stars with temperatures greater than 5000~K. Since a value of $A_V\geq2$ also requires the temperature to be 5000~K or greater, the lack of such spectral features also limits the parameter space for $A_V$. Therefore, we only consider models with effective temperatures under 5000~K and $A_V<2$.}

{While $R_V=5.5$ may be too high for stars in Taurus-Auriga, the interstellar value of $R_V=3.1$ would likely be too low. We measured the effect by using the interstellar value of $R_V=3.1$ to measure $A_V$ and compared those values to the ones determined using $R_V=5.5$. We found that using the interstellar value systematically increased the $A_V$ values by an average of 0.15 mags and the $T_eff$ values increased by an average of 84~K. These shifts are systematic and on the same scale as the uncertainties returned by the fits.}

\section{Composite Models of Spotted Weak-Lined T Tauri Stars} \label{sec:models}
In this section, we first describe our two-temperature fitting routine and apply it to de-reddened stellar spectra to simultaneously determine spot and photospheric parameters ($f_{spot}$, $T_{phot}$, and $T_{spot}$), as well as disk filling factors ($f_{disk}$). {We use $T_{phot}$ to refer to the temperature of the unspotted stellar photosphere, while other studies have referred to this as $T_{q}$ (for quiescent) such as \citet{Fang2016}. $T_{spot}$ in this context refers to the temperature of the spotted regions of the photosphere. These two parameters are identical to those \citet{Gangi} refers to as $T_{hot}$ and $T_{cool}$. Here instead we follow the nomenclature of \citet{myself} for $T_{phot}$ and $T_{spot}$.} 
After describing our fitting routine, we direct our attention to analysing the behavior of our fitting parameters across our multi-epoch spectra spanning an entire stellar rotation and look for evidence of a correlation between photometric and spectroscopic variability. Finally, we quantify the effects of extinction in our fitting routine, as the visual extinctions found in section~\ref{sec:avteff} will directly impact the results of our two-temperature fitting routine.

\subsection{Two-Temperature Fitting}
Recent work by \citet{myself} used Spex spectral library standards \citep{Rayner2009} to construct empirical composite models of spotted stars. Their method used a hotter template to represent the photosphere of the star and a cooler template to represent the spots. The two components were combined by performing a sum weighted by an instantaneous spot filling factor, which represents the fraction of the visible stellar surface covered in spots at the time of observation, following the same process as in \citet{Bary2014,Gully2017}. {Here we have followed a similar approach with two key differences.} 

First, we use the theoretical BTSett-CIFIST set of stellar atmospheres \citep{allard2014} instead of Spex standards. The use of these atmospheric models over the previous grid of IRTF Spectral Library standards allows for several improvements in the fitting process. Given the puffed up nature of TTS atmospheres \citep{Flores2022} and the presence of several gravity-sensitive atomic features in their NIR spectra \citep{Rayner2009}, 
using the BTSettl-CIFIST grid of stellar atmospheres allows us to reduce the effect of these systematic differences between TTS and typical main sequence dwarfs caused by differences in surface gravity. For this reason, we have used the $\log g$~=~3.5 set of models, appropriate for TTS. In addition, adopting the BTSettl-CIFIST stellar atmospheres allow us to expand our fitting range down to 0.7~\micron\ whereas the IRTF Spectral Library standards spectral coverage cut off at 0.8~\micron. This extended coverage includes several spot-sensitive TiO molecular bands, providing additional constraints on the model parameters.

The second update to the empirical composite models was the addition of emission from a disk component, which we modeled as a simple blackbody with a temperature of 1500~K and allowed its filling factor to be a free parameter, given the ability of the disk to subtend a projected area much larger than the stellar surface. 

In this way, our spectral models of spotted stars are constructed using normalized theoretical spectra in the form
\begin{equation}
\begin{split}
    F_{\lambda,\ model} = F_{\lambda}(T_{phot})(1-f_{spot}) \\ +\ F_{\lambda}(T_{spot})f_{spot} C_{spot} + B_{\lambda}(T_{disk})f_{disk} C_{disk}
\end{split}
\end{equation}

\noindent
{where $T_{spot}$ and $f_{spot}$ are the spot temperature and spot filling factor, $T_{phot}$ is the temperature of the photosphere, and $C_{spot}$, and $C_{disk}$ are scaling constants. $C_{spot}$, and $C_{disk}$ are calculated as the ratio at 1.0~\micron\ of blackbody spectra with $T_{phot}$ and $T_{spot}$ or $T_{disk}$, respectively.} {Fitting was performed using an \textit{emcee} routine similar to the one used for the Red-NIR $A_{V}$ fits, allowing for fitting and estimation of uncertainties.} Given the level of complexity of the highly degenerate multi-dimensional parameter space, we initialized the MCMC walkers not clustered around the best fit as is commonly done, but at randomized locations following a uniform distribution across the entire parameter space. This was done to ensure that the parameter space is appropriately explored and that if any walker gets stuck in a localized probability minima, the convergence from the other walkers is still reliable. We model the disk using a simple blackbody curve for $B_{\lambda}(T_{disk})$ with $T_{disk}=1500~K$.


Given the bands of poor atmospheric transmission between J and H and between H and K band, we masked the regions between 1.30-1.45~\micron\ and between 1.80-2.05~\micron, respectively, and did not include them in our fitting or posterior analysis. Likewise, we limit our analysis to the spectral window between 0.75~\micron\ and 2.40~\micron\ for the spectra taken in 2019 and 2022 and to 0.82~\micron\ and 2.40~\micron\ in the 2006 spectra to avoid areas of low signal to noise on the edges of the SpeX detector.

\begin{figure}[!ht]
\centering
\includegraphics[width=1\linewidth]{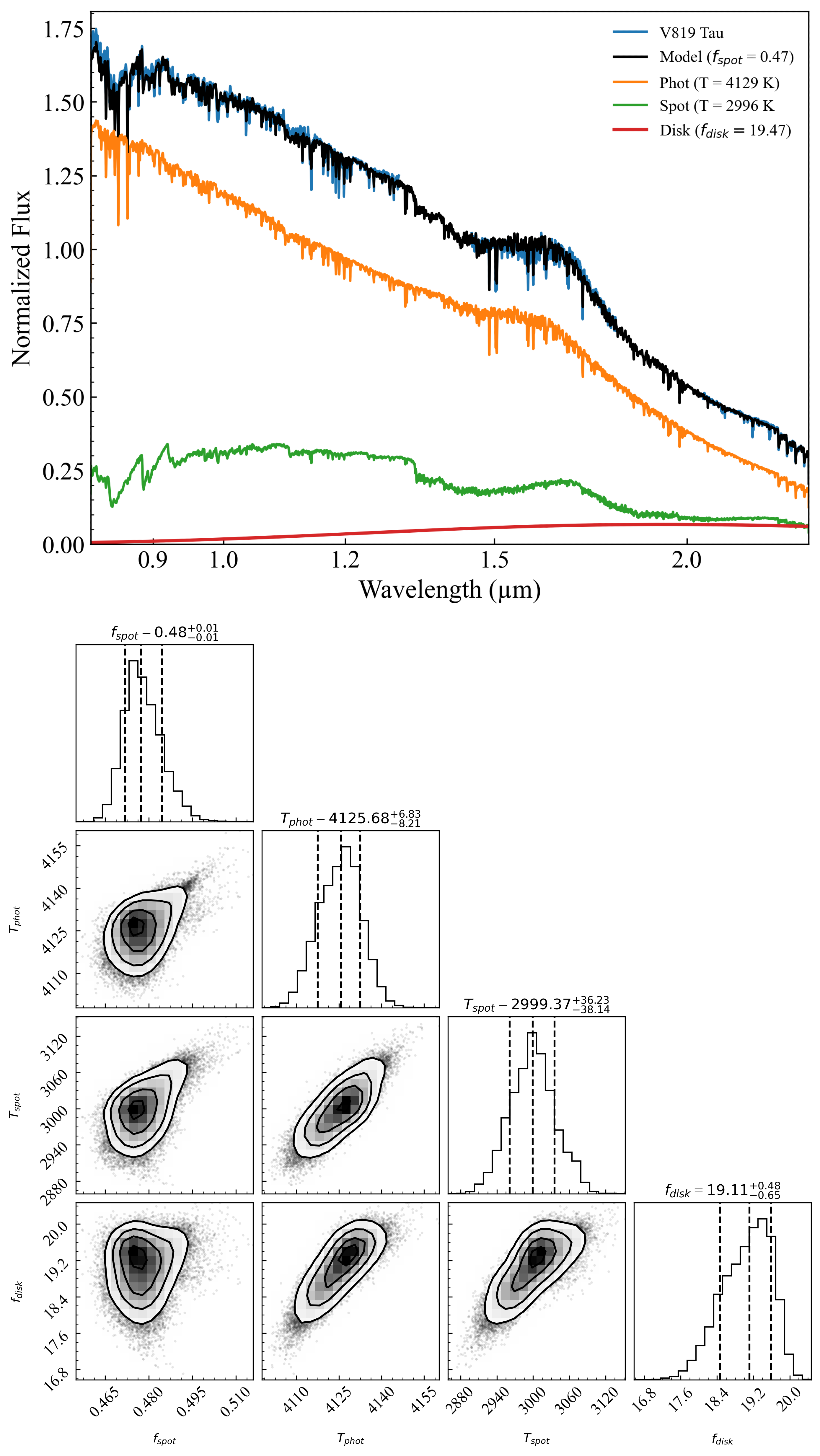}
\caption{Top: A dereddened V819~Tau spectrum (blue) and its corresponding best-fit two-temperature spotted model (black). Plotted are also the photospheric (orange), spot (green), and disk (red) components that make up the fit, weighted by their filling factors so that their sum is the bet-fit model. The spectra have been smoothed for display purposes. Bottom: Corner plot showing the posterior parameter distribution, median, and $1\sigma$ uncertainties for the two-temperature spotted fit.}
\label{fig:v819fit}
\end{figure}

Figure~\ref{fig:v819fit} shows an example of a two-temperature spotted fit for V819~Tau, which is most closely fit by a photosphere at 4126~K covering 52\% of the stellar surface with spots at 3000~K covering the remaining 48\% of the stellar surface. The disk filling factor of V819~Tau is 19.12 times larger than the stellar surface area. This is the largest disk filling factor in our sample of WTTS, with most other fits showing disks filling factors below ten times the visible stellar area. Across the entire spectral range, the two-temperature model successfully reconstructs the SED of V819~Tau as well as the strength of the TiO bands short of 0.9~\micron. Given the temperature of the best-fit photospheric component, these molecular bands, as well as the hump in the H band, are primarily introduced to the model stellar spectrum by the spot component, even though it only contributes a fifth of the stellar flux at the shorter wavelengths. This is particularly significant for the TiO bands in the blue end of the spectra, which are almost entirely derived from the spot component.

Considering the ratio of the photospheric and spot flux, we see that the bluest part of the SpeX spectra will maximize the contribution from the photosphere, while minimizing any contamination from the spots and the disk. This suggests that the 0.75-1.00~\micron\ region used to estimate extinction, while contaminated by spots, will still capture the highest fraction of photospheric flux possible within the SpeX spectra. 

While the model matches the V819~Tau spectrum remarkably well, close inspection reveals that some spectral features are not well-reproduced. The FeH feature at 0.99~\micron, as well as a plethora of atomic lines across the spectrum are deeper in the spectrum than in the model. One possible explanation for these discrepancies lies in our use of $\log g$~=~3.5 stellar atmospheres. Recent work by \citet{LopezValdivia2021} found a mean $\log g$~=~3.87 for TTS in Taurus. Despite these differences, the atomic lines are still more closely matched by the two-temperature fit than by any single-temperature spectral standard. 

\begin{figure*}[!ht]
\centering
\includegraphics[width=1\linewidth]{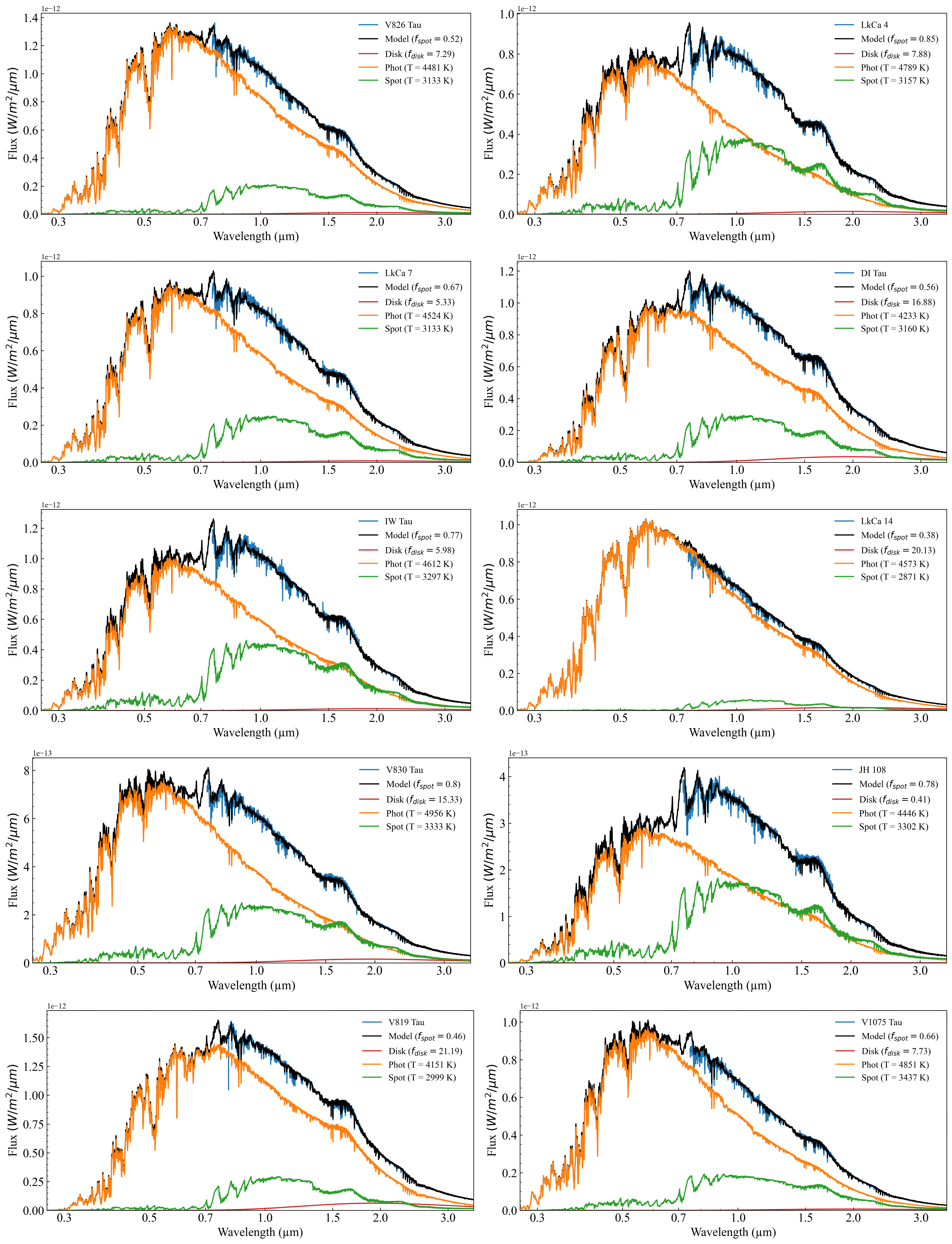}       
\caption{Comparison between best-fit two-temperature composite models (black), and stellar spectra (blue) for our entire sample. Overplotted are the individual photospheric (orange), spot (green), and disk (red) components that make up the model.}
\label{fig:allfits}
\end{figure*}

{As a test of our modeling procedure, we applied the two-temperature photosphere plus cool spot fitting routine to main-sequence spectral standards from the SpeX library.} In these cases, we found that either (1) the models converged on the same temperature for the spots and the photosphere, as seen in the posterior parameter distribution, and the filling factor showed a nonsensical posterior, or (2) the spot filling factor converged to zero or unity, with a photospheric or spot temperature similar to the effective temperature expected from previous studies of the star. Both of these cases represent unspotted stars, and since they were obtained independently by the fitting routine, suggest that our models can rule out unspotted stars. Given that most stars in our sample show evidence for spot modulation in their lightcurves or in the strength of spot-sensitive spectral features, our best-fits suggesting two-temperature models provides strong evidence for the spot interpretation. 

In Figure~\ref{fig:allfits} we present individual fits for all ten stars in our sample following the conventions of Figure~\ref{fig:v819fit}. Similarly to V819~Tau, we find that the spots contribute significantly to the stellar flux across all stars, with JH~108 being the most extreme example with spots contributing as much flux as the photosphere over SpeX wavelengths. {For the three binaries in our sample (V826~Tau, DI~Tau, and IW~Tau) mass ratios close to unity have been found in previous studies \citep{Kraus2011, Reipurth1990}. Therefore, we assume that the spectral emission from both stars is nearly identical. A more thorough discussion of binarity can be found in Section~\ref{sec:binary}.}

\subsection{Fitting Multi-Epoch Spectra of Spotted WTTS}
The large variations seen in the multi-epoch spectra and the corresponding dispersion spectra (See Figure~\ref{fig:spectraobservations}) suggest that the surface starspot configuration of our sample stars undergoes significant evolution on rotational timescales. Under the assumptions of our two temperature model, any difference in the spectrum from night to night is explained as a change in one of the three main parameters: $T_{eff}$, $f_{spot}$, or $T_{phot}$. 

Recent results from \citet{Basri2022} using Kepler lightcurves to estimate spot lifetimes suggest that for stars in the temperature range of our sample stars, spot lifetimes are on the order of 150-200 days. However, their sample did not contain a significant number of young, low-mass stars, instead consisting mostly of main-sequence and sub-giant stars. Since the rotational periods of stars in our sample are all shorter than a week and given the cadence of our spectral acquisition (several spectra per target per night), if the results of \citet{Basri2022} for spot lifetimes extend to younger, less massive TTS, we do not expect to see large differences in spot temperatures across rotational timescales. Examining the two-temperature composite fit of Figure~\ref{fig:v819fit}, the main spectral features distinguishing the photosphere from the spots are the broad TiO absorption bands short of ~0.9~\micron\ and the molecular absorption bands surrounding the H band hump. These features will provide the largest weight in the fit to constrain the spot temperature, and given their strength in typical TTS spectra, variations in their strengths across epochs will be responsible for variations in the best-fit temperatures.

\begin{figure}[!ht]
\centering
\includegraphics[width=1\linewidth]{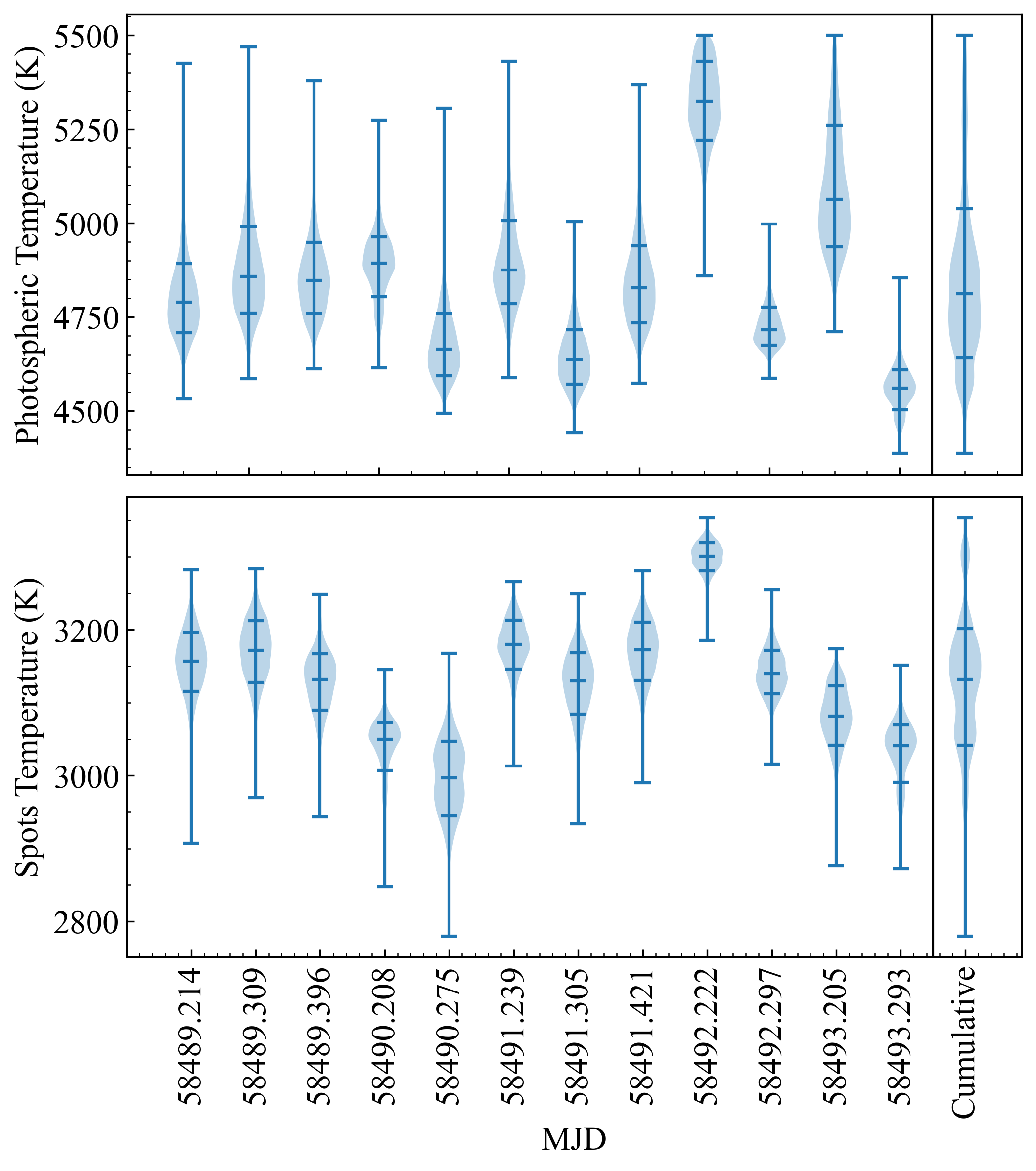}
\caption{Top: Violin plot showing the posterior parameter distribution of $T_{phot}$ across all twelve epochs of observation, as well as the cumulative posterior distribution for LkCa~4. Bottom: Same as above but for $T_{spot}$. Darker blue lines represent the median and $1\sigma$ of each distribution, as well as the extremes for every posterior.}
\label{fig:violin}
\end{figure}

Since we don't limit the parameter space for $T_{phot}$ or $T_{spot}$ during the fitting process or across fits to different epochs, it is possible for best fits for the same star to point to different spot and photospheric temperatures. In practice, we find differences in $T_{phot}$ and $T_{spot}$ on the order of $\approx100$ K across fits to all epochs of available multi-epoch data. We illustrate the scale of these differences in Figure~\ref{fig:violin}, which shows violin plots for the posterior distributions of $T_{phot}$ and $T_{spot}$ across fits to all twelve epochs of LkCa~4 spectra. The last column of both the top and bottom plots of Figure~\ref{fig:violin} shows the \textit{cumulative posterior distribution}, which we construct by stacking the individual posterior distributions for $T_{phot}$ and $T_{spot}$ across all epochs into a single distribution. In the context of typical spot lifetimes, the comparatively short rotational periods of stars in our sample, and the cadence of the observations, these small temperature variations across fits to different epochs are perhaps unsurprising. Owing to these small day-to-day temperature differences, we calculate median values and standard deviations from the cumulative posterior distributions of $T_{phot}$ and $T_{spot}$, and use them as our final photospheric and spot temperatures as well as $1\sigma$ uncertainties. 

The $T_{spot}$, $T_{phot}$, and $T_{spot}$ values returned by the two-temperature fitting routine for every epoch are presented in Table~\ref{tab:appendixtable}, while the final parameters calculated from the cumulative posterior distributions are reported in Table~\ref{tab:spotdata}. For V819~Tau and LkCa~7, for which only single epoch spectra was available, the reported temperatures, filling factors, and uncertainties come directly from the MCMC posteriors to fits to the only available stellar spectrum. For the rest of the stars, photospheric and spot temperatures are calculated as described above, while spot filling factors and uncertainties are the median and standard deviation of filling factors across all epochs. 
\begin{figure*}[!ht]
\centering
\includegraphics[width=1\linewidth]{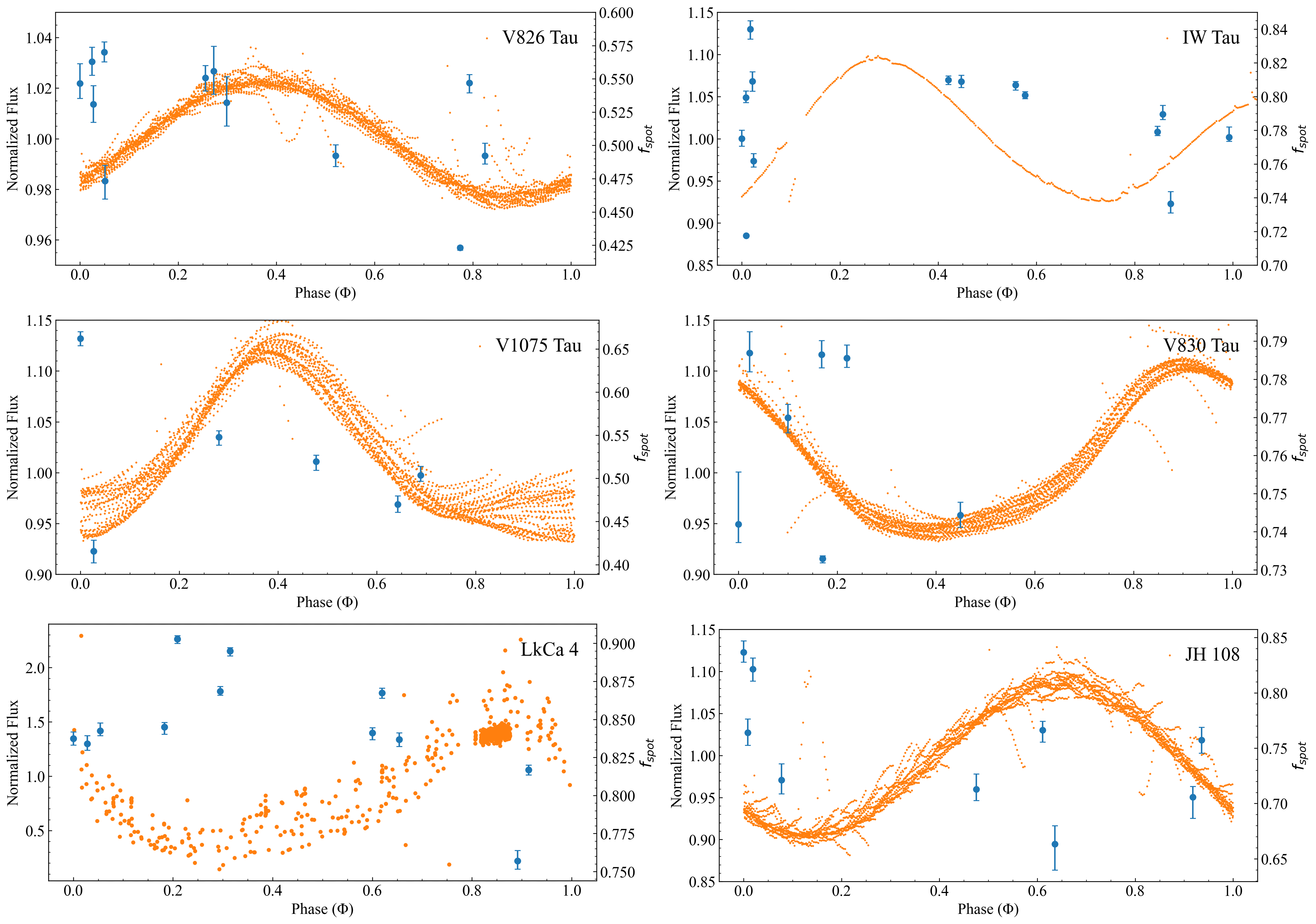}       
\caption{Plotted are the fixed-temperature best-fit filling factors (blue), compared to K2 photometry (orange) as a function of stellar rotation phase for all stars with available photometry, measured rotational periods, and multi-epoch spectra.}
\label{fig:allcurves}
\end{figure*}

\begin{deluxetable*}{c|ccccc|cc} 
\tablecaption{Best-Fit Two-Temperature Spotted Parameters.}\label{tab:spotdata}
\tablehead{
\colhead{Star} & \colhead{$T_{phot}$} & \colhead{$T_{spot}$}  & \colhead{$f_{spot}$} & \colhead{$f_{disk}$} & \colhead{$T_{eff}$} & \colhead{$Log\ (L/L_{\odot})$} & \colhead{$R/R_{\odot}$}\\
\colhead{} & \colhead{[K]} & \colhead{[K]} & \colhead{} & \colhead{} & \colhead{[K]} & \colhead{} & \colhead{}}
\small
\setlength{\tabcolsep}{6pt}
\decimals
\startdata
DI Tau & $4935\pm{183}$ & $3414\pm{39}$ & $0.82\pm{0.04}$ & $12.54\pm{2.01}$ & $3849\pm{51}$ & $-0.15\pm{0.04}$ & $1.89\pm{0.14}$\\ 
IW Tau & $4759\pm{175}$ & $3254\pm{58}$ & $0.79\pm{0.03}$ & $4.15\pm{6.12}$ & $3746\pm{140}$ & $-0.11\pm{0.07}$ & $2.07\pm{0.19}$\\
JH 108 & $4624\pm{201}$ & $3402\pm{29}$ & $0.76\pm{0.05}$ & $14.71\pm{10.36}$ & $3831\pm{104}$ & $-0.36\pm{0.08}$ & $1.48\pm{0.09}$\\
LkCa 4 & $4755\pm{119}$ & $3149\pm{67}$ & $0.84\pm{0.04}$ & $6.52\pm{2.10}$ & $3557\pm{99}$ & $-0.21\pm{0.11}$ & $2.06\pm{0.30}$\\
LkCa 7 & $4524\pm{54}$ & $3148\pm{60}$ & $0.67\pm{0.03}$ & $5.33\pm{0.76}$ & $3779\pm{85}$ & $-0.23\pm{0.06}$ & $1.80\pm{0.13}$\\
LkCa 14 & $4573\pm{30}$ & $2827\pm{88}$ & $0.42\pm{0.03}$ & $8.52\pm{5.48}$ & $4123\pm{49}$ & $-0.21\pm{0.06}$ & $1.58\pm{0.10}$\\
V819 Tau & $4151\pm{8}$ & $2998\pm{35}$ & $0.46\pm{0.01}$ & $21.19\pm{0.72}$ & $3753\pm{85}$ & $0.02\pm{0.06}$ & $2.42\pm{0.13}$\\
V826 Tau & $4492\pm{92}$ & $3138\pm{69}$ & $0.54\pm{0.04}$ & $12.99\pm{3.58}$ & $3996\pm{42}$ & $-0.01\pm{0.04}$ & $2.04\pm{0.09}$\\
V830 Tau & $4894\pm{96}$ & $3398\pm{46}$ & $0.77\pm{0.02}$ & $10.16\pm{3.81}$ & $3917\pm{37}$ & $-0.26\pm{0.04}$ & $1.63\pm{0.07}$\\
V1075 Tau & $4623\pm{121}$ & $3168\pm{207}$ & $0.51\pm{0.08}$ & $8.44\pm{4.66}$ & $4118\pm{99}$ & $-0.20\pm{0.04}$ & $1.52\pm{0.08}$\\
\enddata
\end{deluxetable*}


\subsection{Correlating Photometric Variability with Spectroscopic Variability}
Our current understanding of T~Tauri star variability suggests that starspots rotating into and out of view cause the star to appear dimmer and brighter due to their lower luminosity compared to the stellar photosphere \citep[e.g.,][]{Vrba1986, Herbst1994}. The photosphere-to-spot ratio, median filling factor, and amplitude with which the brightness varies, modulated by the stellar rotation period, all determine the behavior of the stellar lightcurve. If the spectral variability is large enough, the lightcurve should be negatively correlated with the filling factors as a function of phase with epochs of photometric maxima corresponding to spot minima \citep{myself}. 

To search for such a correlation, we used the fitting described in Section~\ref{sec:models} but modified it to fix the spot and photospheric temperatures to the median values of the best-fits across all epochs (See Table~\ref{tab:spotdata}). We then re-fit every epoch of observation, allowing only $f_{spot}$ to vary. By fixing $T_{phot}$ and $T_{spot}$, any spectral variability is correlated with a difference in filling factors across observations. Otherwise with all parameters allowed to vary freely the same filling factor with different values of $T_{phot}$ and $T_{spot}$ could reproduce the same stellar spectrum and lead to a degeneracy across models. 

In Figure~\ref{fig:allcurves}, we plot photometry along with the best-fit filling factors as a function of phase for all stars with available photometry and well-determined rotation rates. We use K2 photometry for all stars except LkCa~4, for which we use AAVSO V-band photometry. The data {were} phased by setting the first observation as $\Phi=0$ and using the rotation periods from Table~\ref{tab:obsdata}. For IW~Tau, which possesses two accepted periods of 5.50 and 7.04 days, we used the longer of the two periods and show photometry for one stellar rotation only.

LkCa~4 and JH~108 show filling factors that are at a maximum when the lightcurve is at a minima, and photospheric maxima that correspond to lowest spot coverage. This is what we would expect to see if the variability was caused solely by spots rotating into and out of view. Given what we know about spot lifetimes, the relatively short rotation periods for these stars and the timeline of our observations, we expect the correlation seen in JH~108 and LkCa~4 to be real and not caused by changes in the overall spot coverage. Interpreting results for V826~Tau, IW~Tau, V1075~Tau, and V830~Tau is more complicated, with no evident signs of a correlation with phase. Given the lower amplitude of the variability observed in their lightcurves, it is possible that the night-to-night variations in the spectra are too small to lead to an observable correlation in filling factors. This is particularly true when compared to LkCa~4, which exhibits variability {of} over half a magnitude.

\subsection{The Effects of Extinction on Spot Fitting Results}
\begin{figure}[!ht]
\centering
\includegraphics[width=1\linewidth]{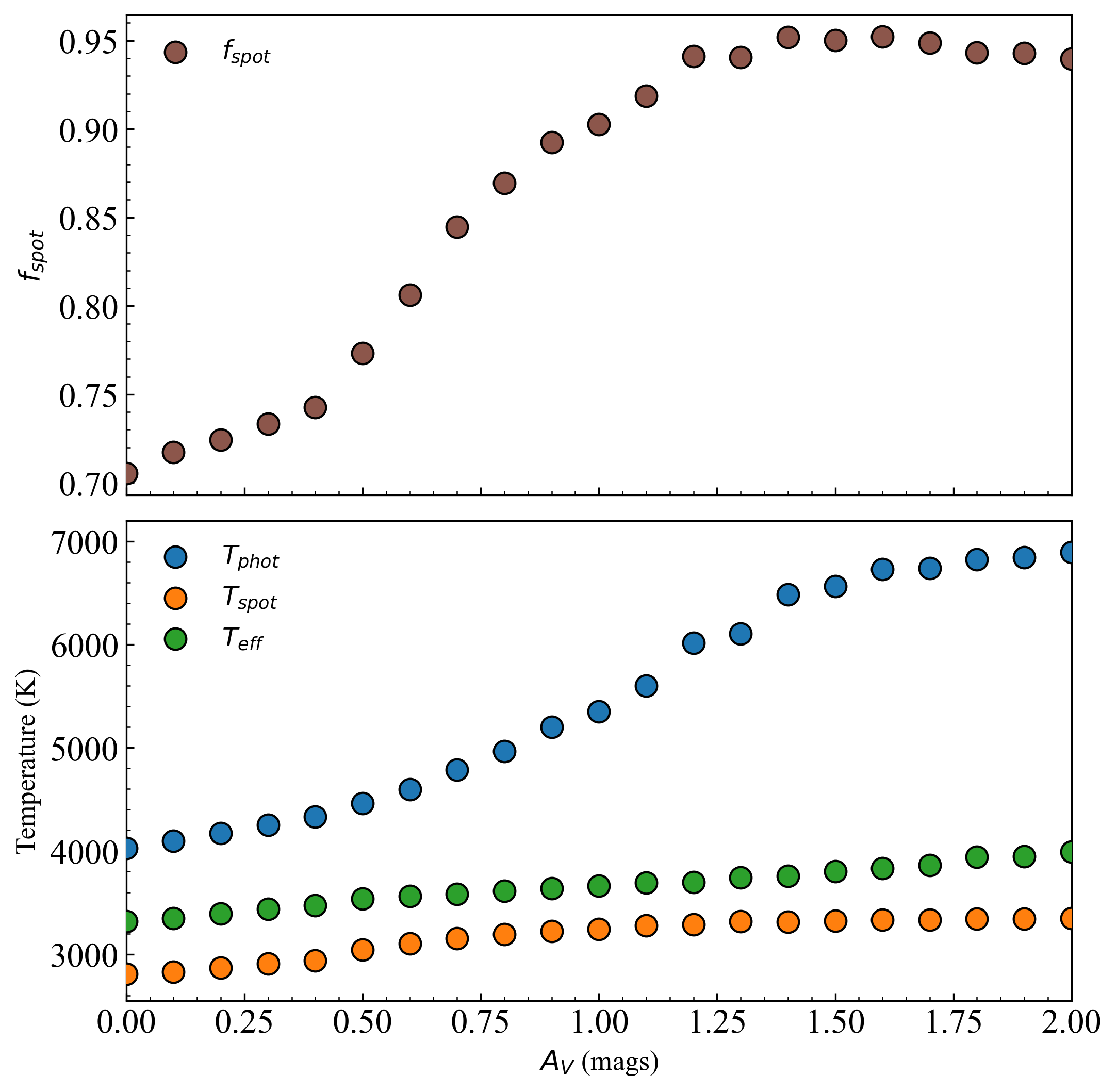}       
\caption{Top: Resulting best-fit filling factors (brown) from fitting a single LkCa~4 spectrum dereddened in 0.1 mag increments between zero and two magnitudes of extinction. Bottom: Same as above, but plotted are the corresponding best fit photospheric (blue) and spot (orange) temperatures, as well as the resulting effective temperature (green) for each combination of $T_{phot}$, $T_{spot}$, and $f_{spot}$ in the fits.}
\label{fig:avspotfit}
\end{figure}

In section~\ref{sec:avteff}, we determined optical extinctions for all targets in our sample and then de-reddened the spectra accordingly prior to applying our two-temperature fitting routine. Since the shape of the continuum of the stellar spectrum plays a role in constraining spot and photospheric parameters, the choice of $A_{V}$ will affect the best-fit parameters. To quantify the effect of extinction on the resulting best-fit $T_{phot}$, $T_{spot}$, and $f_{spot}$ values, we dereddened a LkCa~4 spectrum between zero and two magnitudes of visual extinction using steps of 0.1 magnitudes, the same extinction law and $R_{V}$ value described in Section~\ref{sec:avteff}, and then applied our two-temperature composite fitting routine. Figure~\ref{fig:avspotfit} shows variations in $T_{phot}$, $T_{spot}$, and $f_{spot}$ as a function of $A_{V}$. For low values of $A_{V}$, the strengths of the molecular bands in the I, Y, and H bands provide strong constraints on the spot temperature, while the slope of the spectrum provides constraints on the photospheric temperature. 

As extinction increases, the strength of the molecular features remains constant and so the spot temperature shows a slight increase while the photospheric temperature increases almost linearly with extinction inn order to compensate for the steeper continua. As extinction surpasses $A_{V}=1.7$ mags, the photospheric component cannot increase any further due to the upper limit on the temperatures for the BTSettl-CIFIST atmospheres. Given that several strong temperature-sensitive molecular features only exist in the spot spectrum, the choice of extinction will only impact the photospheric temperature, as the spot temperature will be determined by the requirement that the molecular features be present in the best-fit model. As extinction increases, several high-level and ionization atomic lines appear in the photospheric spectrum that are not present in the LkCa~4 spectrum, and that could not be fully obscured due to high veiling. While the photospheric temperature continues to increase, the molecular features remain, and since they can only be introduced to the model from the spot spectrum, the spot temperatures only show slight variations.

In summary, two-temperature fitting is only sensitive to extinction insofar as it will be correlated with the photospheric temperature, while spot temperatures can be readily determined even when extinctions have been poorly determined. For every magnitude increase in $A_{V}$, the photospheric temperature of our models increases by nearly two-thousand degrees, while the spot temperatures increase by a couple hundred of degrees. Given the high spot filling factors, a modest temperature increase in the spot temperature will lead to a similarly sized increase in the effective temperature, whereas even significant differences in photospheric temperatures will have a much smaller effect.
{Given the results shown in Figure~\ref{fig:avfit}, our use of $R_V=5.5$ leads to a systematic underestimation of the T$_{phot}$ and T$_{spot}$ by an average of $\sim$100~K and $\sim$50~K, respectively.}

\section{The Challenges of Placing Spotted Stars on an HR Diagram}\label{sec:HR}
In order to place stars on an HR Diagram, we must determine both effective temperatures and stellar luminosities. In this section, we derive these parameters for three different relevant cases using values for $T_{eff}$ and $L_{\star}$ obtained from: 1) optical studies in the literature, 2) our Red-NIR fits, and 3) the photosphere+spots multi-component fitting. With these values calculated, we can then place the stars in our sample on an HR diagram alongside unspotted and spotted evolutionary models, using them to infer new ages and masses and discuss our results.

\subsection{Optical and Red-NIR Temperatures and Luminosities}
We adopt optical spectral types and temperatures for stars in our sample from \citet{Dent2013} and \citet{andrews2013} as specified in Table~\ref{tab:obsdata}. For \citet{Dent2013}, we adopt their spectral types and match them to optical temperatures using the conversions of \citet{Pecaut2013} to derive final $T_{opt}$ values. To calculate stellar luminosities, we match a normalized synthetic BTSettl-CIFIST atmosphere of corresponding temperature ($T_{opt}$) to the approximately flux calibrated {and dereddened} SpeX spectra before integrating between 300 nm and 75~\micron\ to calculate the bolometric stellar fluxes. Finally, we use Gaia EDR3 distances to calculate the stellar bolometric luminosity $L_{opt}$. Since \citet{andrews2013} provide luminosities on the basis of integrated optical-NIR SEDs, no further work is needed and we merely adopt their values. We repeat this analysis using the Red-NIR ($T_{red}$) temperatures found in the $A_{V}$ fits (See Section~\ref{sec:avteff}) to derive Red-NIR luminosities. The corresponding temperatures and luminosities for both optical and Red-NIR regimes are shown in Table~\ref{tab:obsdata} for all targets.

\subsection{Spotted Temperatures and Luminosities}
Using our constraints on $T_{phot}$, $T_{spot}$, and $f_{spot}$ we calculate effective temperatures following \citet{Gully2017} as
\begin{equation}
    T_{eff} = \left[T_{phot}^4(1-f_{spot})+T^4_{spot}f_{spot}\right]^{1/4}
\end{equation}
\noindent the resulting effective temperatures are reported in column six of Table~\ref{tab:spotdata}, where the values presented stem from the median $T_{eff}$ across all epochs. The reported uncertainties correspond to 1$\sigma$ standard deviations. We report these based on the fact that the dispersion in $T_{phot}$, $T_{spot}$, $f_{spot}$, and the corresponding $T_{eff}$ across all epochs {exceeds} the individual uncertainties in the parameters corresponding to individual epochs. For stars with only single-epoch data, we use the median value for the uncertainties in the other stars with multi-epoch data under the assumption that the dispersion seen in our multi-epoch sample is characteristic of the typical dispersion for stars of this type. Given the high filling factors for stars in our sample, we find that effective temperatures match the best-fit spot temperatures better than the best-fit photospheric temperatures. 

With photosphere+spot effective temperatures calculated, we repeated the process outlined for the single-temperature optical and Red-NIR cases with one small difference: the non-stellar flux contribution from the disk was subtracted from the two-temperature spotted models before using Gaia EDR3 distances to calculate the bolometric stellar luminosity. These parameters are presented in Table~\ref{tab:spotdata}. We present these stellar radii and luminosities with the caveat that our spectra have only been approximately flux-calibrated using A0V standard stars as part of the SpeX reduction package. For this reason, we expect there to be a systematic uncertainty in our measured radii. However, these should be mitigated by averaging across all of our observations taken on different nights and with multiple A0V standards.

Extended models spanning the entire 300~nm -- 75~$\micron$ spectral range and flux-matched to the observed stellar spectrum are presented in Figure~\ref{fig:allfits} for all stars and highlight several important conclusions. The ratio of the photospheric flux to that of the spots flux ranges between 20 and 40 over optical wavelengths. The ratio becomes significantly smaller for longer wavelengths. Given the overwhelming contribution of the photosphere at optical wavelengths, spectral classifications using optical spectra will have been more sensitive to the photosphere of the star, where those using spectra at longer wavelengths become increasingly affected by starspot emission. In this context starspots arise as a compelling explanation for the mismatches between optical and infrared spectral types. These results, when combined with those of \citet{myself} provide substantial evidence for starspots being responsible for both the mismatches previously mentioned as well as for {some} the photometric variability characteristic of both WTTS and CTTS.

\subsection{Calculating Optical, Red-NIR, and Photosphere+Spot Ages and Masses}
Recent evolutionary models from \citet{Somers2020} account for the impact of starspots on the stellar interiors, taking into account the flux-blocking effect of starspots, redirected energy flow, and modified surface temperature boundary conditions. Evolutionary tracks and isochrones for spotted stars are computed in steps of 0.05~$M_{\odot}$ for masses between 0.1~$M_{\odot}$-- 1.3~$M_{\odot}$ and for filling factors of $f_{spot}$~=~0,~0.17,~0.34,~0.51,~0.68,~and~0.85. In particular, the $f_{spot}=0$ set of tracks represent unspotted stars, and in practice show very small deviations from traditional unspotted models such as those of \citet{Baraffe2015}.

\begin{figure}[!h]
\centering
\includegraphics[width=1\linewidth]{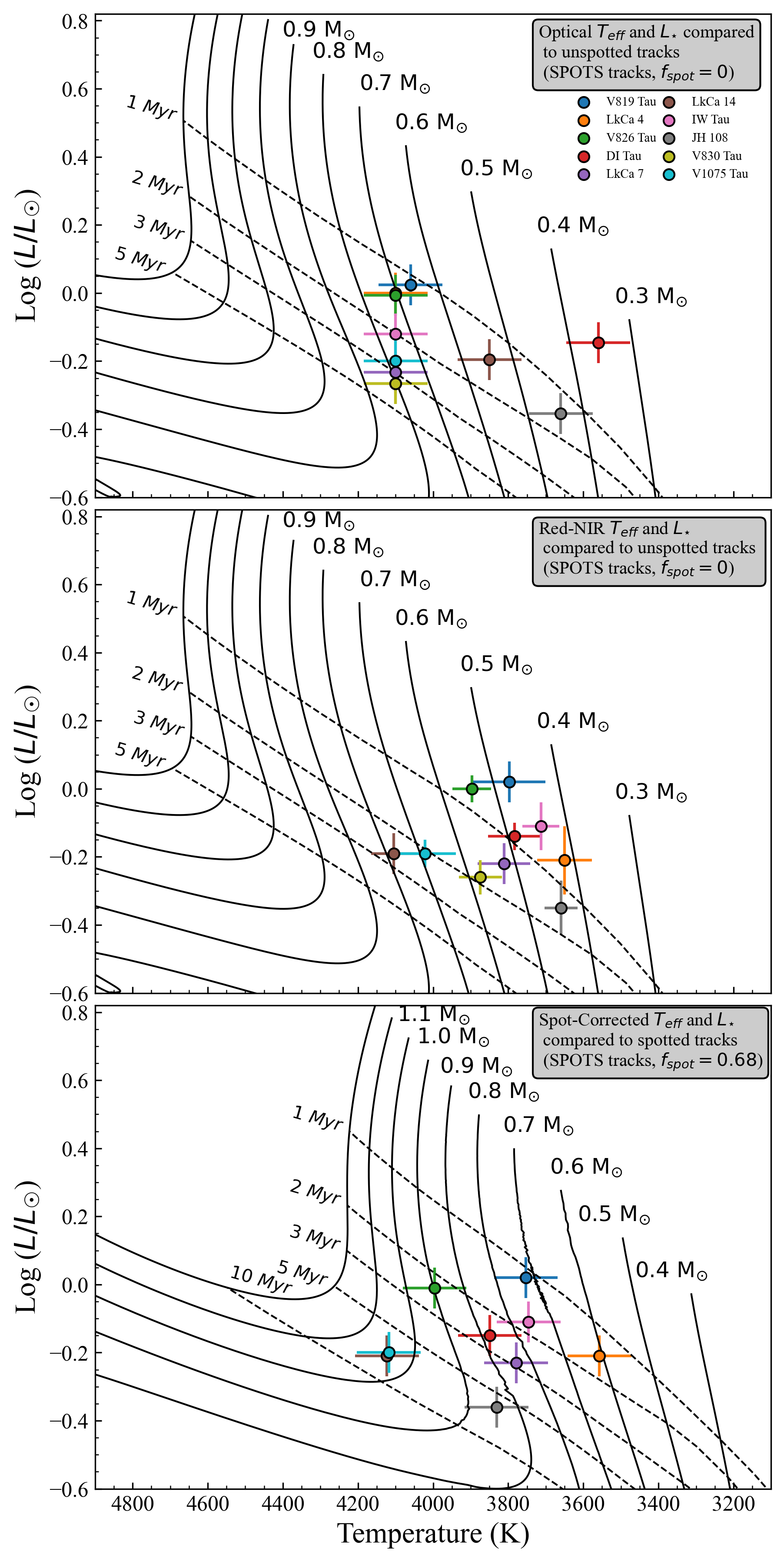}
\caption{Top and Middle: HR Diagrams comparing optical and Red-NIR temperatures and luminosities, respectively, for WTTS in Taurus-Auriga to unspotted ($f_{spot}=0$) evolutionary models. Bottom: Plotted is the same comparison as before, but using spot-corrected temperatures and luminosities and spotted ($f_{spot}=0.68$) evolutionary models. Each color corresponds to the same star across subplots.}
\label{fig:hrcomparison}
\end{figure}

In the top panel of Figure~\ref{fig:hrcomparison} we place stars on an HR Diagram according to their optical spectral types, using the values for $T_{opt}$ and $L_{opt}$ we calculated in the previous section. Alongside these, we plot unspotted ($f_{spot}=0$) evolutionary tracks from the SPOTS models in $0.1M_{\odot}$ increments as well as appropriate isochrones. To calculate ages and masses, we perform a linear interpolation across the set of evolutionary tracks and then use our best-fit median $T_{eff}$ and $L_{\odot}$ to infer ages and masses. This is performed across the whole 0.05 $M_{\odot}$ spaced grid to reduce any error caused by spacing between tracks. This is particularly important for the region of the tracks where stars take a sharp turn to hotter temperatures around 10~Myr. Optical ages and masses are shown in Table~\ref{tab:masses}. 

We repeat this process for the best-fit Red-NIR temperatures ($T_{red}$) and luminosities ($L_{red}$) as shown in the middle panel of Figure~\ref{fig:hrcomparison} to derive Red-NIR ages and masses. These are also presented in Table~\ref{tab:masses}. Comparing optical and Red-NIR HR Diagram locations, we can see that stars are, on average, shifted to cooler temperatures while luminosity remains roughly constant. This has the effect of lowering the inferred Red-NIR masses with respect to optical masses. The median mass drop corresponds to $0.2M_{\odot}$, while age decreases by a median 0.29 Myr. 

To calculate ages and masses using our photosphere+spot temperatures $T_{spot}$ and luminosities $L_{spot}$ we must use evolutionary tracks calculated for the corresponding stellar value of $f_{spot}$. Because of this, we perform the same interpolation across evolutionary models as before with the added dimension of filling factor, allowing us to calculate a set of tracks for any given value of $f_{spot}$. Ages and masses derived this way are reported in Table~\ref{tab:masses}. 

\begin{figure}[!h]
\centering
\includegraphics[width=1\linewidth]{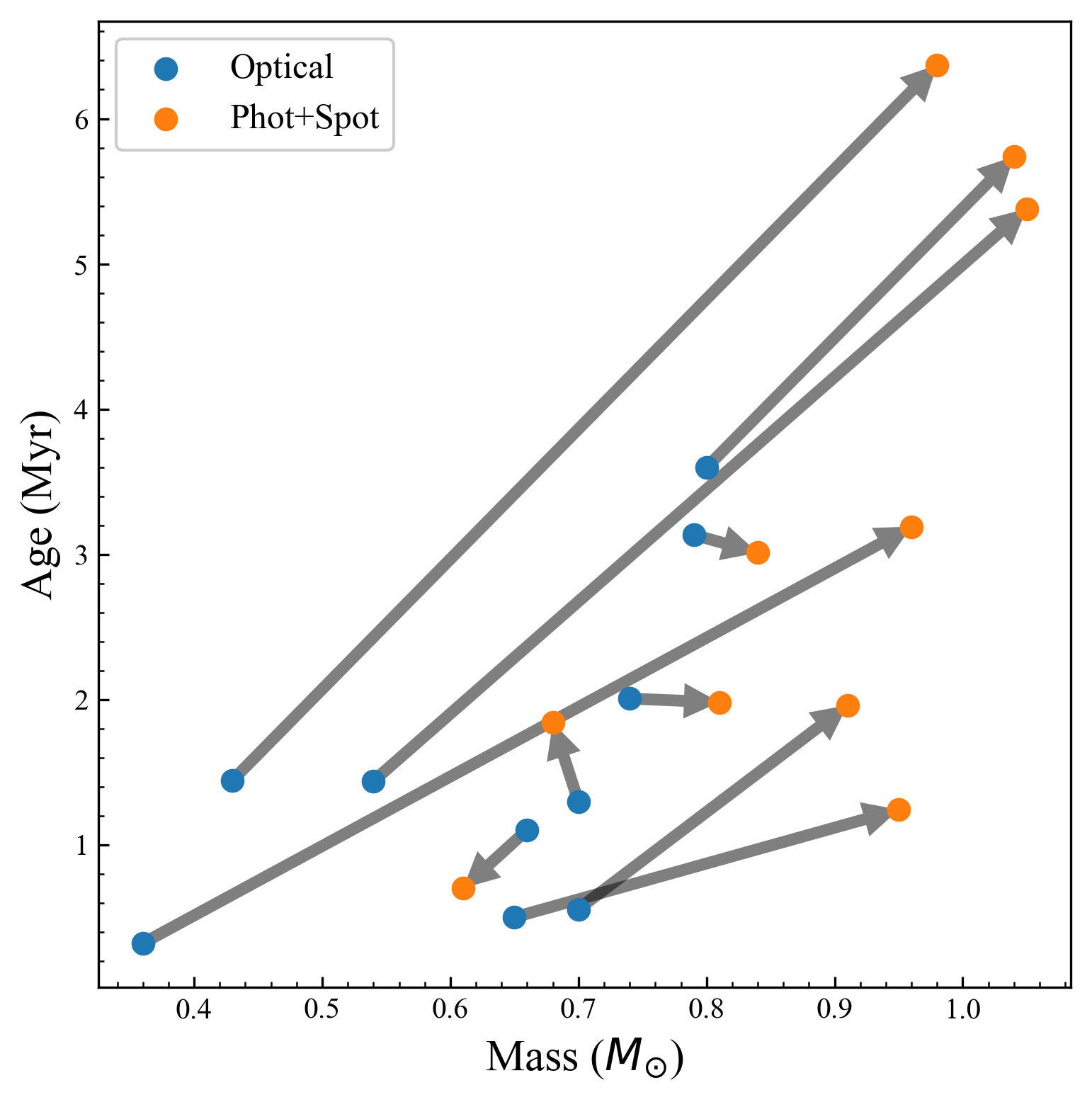}
\caption{Scatter plot showcasing mass and age shifts caused by spots. While a couple of stars decrease in age and mass when spots are taken into account, the overall effect is to significantly increase ages and masses.}
\label{fig:scatter}
\end{figure}

The bottom panel of Figure~\ref{fig:hrcomparison} shows our sample stars alongside a set of $f_{spot}=0.68$ spotted evolutionary tracks and isochrones. Since each star has its own filling factor that needs to be used when computing spot-corrected ages and masses, we decided to plot $f_{spot}=0.68$ merely as an instructional example, as it is close to the median $f_{spot}$ across the entire sample. We find two notable differences between the optically-derived placements on the HR Diagram and those derived from Red-NIR spectral region. Due to the lower effective temperatures returned by the photosphere+spot fits, stars are shifted to lower temperatures to the right of the HR Diagram, with luminosities that remained roughly constant. While this is similar to the shifts seen between optical and Red-NIR locations, there is an added shift in the tracks themselves whereby tracks for any given mass are shifted to lower temperatures. Due to the convection blocking effect of starspots, the outward transport of energy within spotted stars is suppressed, causing an additional shift in the isochrones to higher luminosities. These two effects (cooler effective temperatures and shifted tracks) when combined have the net effect of shifting stars to higher masses and ages when compared to both optically and Red-NIR derived values even though the effective temperatures themselves are lower. The magnitude of these shifts is shown in Figure~\ref{fig:scatter}. It is notable that while a couple of stars do become younger or less massive, the most typical shift is towards higher values. 

\begin{deluxetable*}{c|cc|cc|ccc}
\tablecaption{Optical, Red-NIR, and Spot-Corrected Stellar Ages and Masses.}\label{tab:masses}
\tablehead{
\colhead{Star} & \colhead{$M_{\mathrm{opt}}/M_{\odot}$} & \colhead{$Age_{\mathrm{opt}}$ [Myr]}  & \colhead{$M_{\mathrm{red}}/M_{\odot}$} & \colhead{$Age_{\mathrm{red}}$ [Myr]} & \colhead{$M_{\mathrm{spot}}/M_{\odot}$} & \colhead{$Age_{\mathrm{spot}}$ [Myr]}\\
\colhead{} & \colhead{} & \colhead{} & \colhead{} & \colhead{} & \colhead{} & \colhead{}}
\decimals
\startdata
DI Tau & 0.36 & 0.32 & $0.49^{+0.06}_{-0.05}$ & $0.86^{+0.56}_{-0.54}$ & $0.96^{+0.07}_{-0.05}$ & $3.19^{+0.72}_{-0.71}$ \\
IW Tau & 0.74 & 2.01 & $0.44^{+0.05}_{-0.05}$ & $0.51^{+0.31}_{-0.24}$ & $0.81^{+0.16}_{-0.14}$ & $1.98^{+1.04}_{-0.79}$\\
JH 108 & 0.43 & 1.44 & $0.43^{+0.06}_{-0.05}$ & $1.50^{+0.55}_{-0.79}$ & $0.98^{+0.06}_{-0.09}$ & $6.37^{+1.66}_{-3.25}$\\
LkCa 4 & 0.70 & 1.30 & $0.42^{+0.05}_{-0.05}$ & $0.62^{+0.50}_{-0.36}$ & $0.68^{+0.10}_{-0.09}$ & $1.84^{+0.80}_{-1.02}$\\
LkCa 7 & 0.79 & 3.13 & $0.52^{+0.06}_{-0.06}$ & $1.44^{+0.41}_{-0.49}$ & $0.84^{+0.09}_{-0.10}$ & $3.01^{+1.13}_{-1.60}$\\
LkCa 14 & 0.54 & 1.44 & $0.78^{+0.11}_{-0.10}$ & $2.66^{+0.98}_{-0.93}$ & $1.05^{+0.04}_{-0.04}$ & $5.38^{+1.32}_{-1.66}$\\
V819 Tau & 0.66 & 1.10 & $0.47^{+0.05}_{-0.05}$ & $0.37^{+0.17}_{-0.14}$ & $0.61^{+0.08}_{-0.07}$ & $0.70^{+0.48}_{-0.44}$\\
V826 Tau & 0.70 & 0.56 & $0.54^{+0.06}_{-0.05}$ & $0.63^{+0.42}_{-0.40}$ & $0.91^{+0.05}_{-0.05}$ & $1.96^{+0.35}_{-0.37}$\\
V830 Tau & 0.80 & 3.60 & $0.58^{+0.08}_{-0.07}$ & $1.86^{+0.53}_{-0.52}$ & $1.04^{+0.02}_{-0.04}$ & $5.74^{+1.04}_{-1.12}$\\
V1075 Tau & 0.78 & 2.74 & $0.69^{+0.10}_{-0.08}$ & $2.08^{+0.71}_{-0.64}$ & $1.08^{+0.05}_{-0.09}$ & $5.62^{+1.95}_{-1.39}$\\
\hline
Median $\pm\ \sigma$ & $0.70\pm0.16$ & $1.44\pm1.09$ & $0.50\pm0.12$ & $1.15\pm0.77$ & $0.94\pm0.16$ & $3.10\pm2.02$\\
\enddata
\end{deluxetable*}

\begin{figure*}
\centering
\includegraphics[width=1\linewidth]{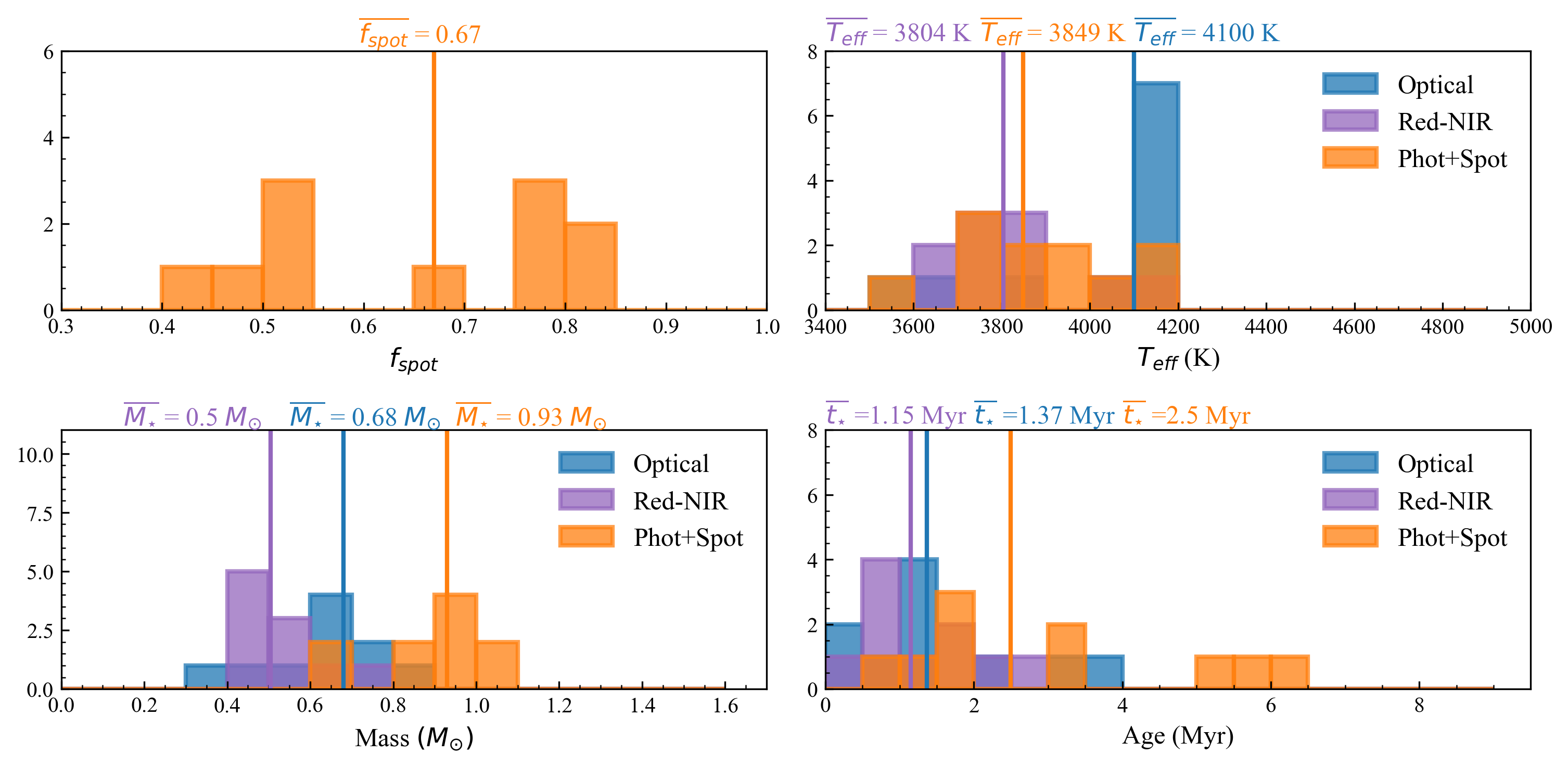}
\caption{Top Left: Histogram showing filling factors ($f_{spot}$) for stars in our sample. Top Right: Histogram showing optical (blue), Red-NIR (purple) and spot-corrected (orange) effective temperatures ($T_{\star}$). Bottom Left: Histogram showing inferred masses from the three previously mentioned effective temperatures ($M_{\star}$). Bottom Right: Same as before but for ages ($t_{\star}$). Overplotted are median values for all histogram parameters ($\overline{f_{spot}}$, $\overline{T_{eff}}$, $\overline{M_{\star}}$, and $\overline{t_{\star}}$).}
\label{fig:masscomparison}
\end{figure*}

Comparing Red-NIR and photosphere+spot HR Diagram placements, we see that temperature and luminosity differences are small in comparison to those between optical and photosphere+spot. Therefore, the main source of discrepancy between Red-NIR and photosphere+spot masses stems from the shifts in the location of the evolutionary tracks and isochrones for any given mass and age. These comparisons are made clearer in Figure~\ref{fig:masscomparison}, which shows histograms for $f_{spot}$, $T_{eff}$, $M_{\star}$, and Age for the three different scenarios used: Optical, Red-NIR and photosphere+spot fitting. Overplotted are median values for every quantity, highlighting for instance the small difference in median effective temperatures (3804~K vs 3849~K) seen between Red-NIR and photosphere+spot temperatures and larger differences with optical (4100~K) temperatures. These temperature shifts, in conjunction with shifts in the location of the evolutionary tracks, result in the apparent mass spreads seen where Red-NIR masses are underestimated with respect to optical and photosphere+spot masses. Optical masses are higher than Red-NIR masses, but still don't agree with the masses obtained from photosphere+spot fitting because the unspotted $f_{spot}=0$ tracks do not see temperature shifts resulting from the mechanical effect of spots. Therefore, photosphere+spot masses are higher than both of these on average. We see a similar situation with ages where optical and Red-NIR ages are in close agreement with median values of 1.37 and 1.15 Myr, respectively, while spotted ages have a median value of 2.5 Myr and, more importantly, span a much larger range of values.   

Placing spotted PMS in an HR Diagram correctly is complicated. The use of Red-NIR temperatures provides a more accurate representation of the surface effective temperature than optical temperatures, but comparing these lower effective temperatures to traditional unspotted evolutionary models results in a larger underestimation of the stellar mass when compared to optical values. For the shifts between optical and photosphere+spot ages and masses, the magnitude of the effect depends on the specific $T_{phot}$, $T_{phot}$, and $f_{spot}$ of the star, but we find a mean shift of $\Delta M~=~0.24\ M_{\odot}$, corresponding to a 35\% average increase in mass. The resulting shifts in age are on average $\Delta Age = 1.7$~Myr, corresponding to a doubling in age. These results highlight the significant impact that starspots can have on age and mass determinations for the youngest stars. Such shifts are likely more significant for lower mass stars \citep[e.g.,][]{Flores2022} with higher spot coverage. The effects of starspots extend further into radii, where we find systematically larger radii than those predicted by stellar evolutionary models that ignore the effects of spots \citep[e.g., ][]{Baraffe2015} for stars of the same age and mass.

These results are significant, but unsurprising when considered in conjunction with the observed spectral type mismatches and mass and age spreads seen in star forming regions. The ability for starspot spectral models to simultaneously reproduce both of these effects by modifying stellar spectra and inferred ages and masses represents strong evidence for the conclusion that starspots are to blame for these effects. 

As was previously stated, \citet{Fischer2011} showed that the region between $0.7-1.0\ \micron$ is the least affected by veiling, making it ideal for studying young stellar systems often embedded within molecular clouds and disks. However, this region also shows the most contamination from starspots, resulting in large spectral type biases that lead to further biases in the inferred masses and ages. Our results showcase that the presence of starspots needs to be properly accounted for in order to measure veiling and extinction accurately in both WTTS and CTTS. 

\section{Discussion}\label{sec:discussion}
\subsection{Are Extreme Starspot Filling Factors Feasible?}
Sunspots can cover just under a percent of the Sun's surface at solar maximum \citep{Penza2021}. By comparison, our results showcasing median starspot filling factors as high as 84\% of the visible stellar surface are difficult to reconcile in comparison to the solar analog. However, they are not atypical to those found in similar studies of young stars. Recent work by \citet{Cao2022} measured starspot filling factors for a sample of 240 stars in the Pleiades and M67, finding median starspot filling factors of $f_{spot}$=0.248 and $f_{spot}$=0.03 respectively. While these results are systematically lower than our values, they correspond to much older stellar populations (Pleiades~$\approx$~125~Myr, M67~$\approx$~4~Gyr). Since starspot coverage is expected to decrease with age \citep{morris2020}, these results are unsurprising.

Similar studies for stars in younger star forming regions find comparably extreme starspot filling factors. \citet{Gully2017} performs two-temperature fits of LkCa~4 finding comparable spot and photospheric temperatures, with a best-fit filling factor of 80\%. These results are supported by those of \citet{myself} who found a median filling factor of 89\% on the basis of both broadband spectral fitting and fitting of strong spot-sensitive TiO features. \citet{Gangi}, whose sample of sixteen spotted stars includes many of the targets we studied, found evidence for starspots covering between 24\% and 90\% of the stellar surface.

\citet{Grankin2008} use UBVRI photometry of a sample of TTS which includes LkCa~4, LkCa~7, LkCa~14, V819~Tau, V826~Tau, and V830~Tau from our sample to estimate the minimum and maximum spot filling factors. In addition, they calculated the temperature differences between the photosphere and the spots required to generate the amplitudes of the variability. Their results highlight the need for spots 500-1400~K colder than the photosphere covering 17-73\% of the stellar surface. Therefore, while our spot filling factors are high, they are predicted by the the large amplitudes observed in the lightcurves of these stars.

The need for high filling factors could also be explained by the inclination of the star, whereby the presence of large polar spots on stars with high inclinations could result in large filling factors, as the only observable region of the star would have a locally higher filling factor than the rest of the stellar surface. We can only probe the emission from the visible regions of the star, and as such neglect any inclination effects in our analysis. Assuming that starspots behave like sunspots and are concentrated near the solar equator, this effect should be small, however, Zeeman-Doppler Imaging of WTTS has shown the presence of strong localized magnetic fields \citep{Donati2023} as well as localized dimming \citep{Donati2017, Yu2019, Donati2019} consistent with starspots on or around the poles. Polar spots, if present, could account for the high filling factors while generating both high and low levels of photometric variability. If starspots are concentrated around the pole and do not go out of view as the star rotates, we would infer an unrealistically high filling factor for the stellar surface while observing a relatively small level of photometric variability. On the other hand, if a star has a larger overall spot coverage but the spots are not concentrated around the pole, it could exhibit similar levels of variability, albeit with a smaller measurable spot filling factor. 

The simple two-temperature stellar atmospheric models could be responsible for the large spot filling factors. Similarly to the sun, we expect to see starspots that lack perfect borders representing large temperature discontinuities, but instead possess extended temperature gradients between the warm surrounding photosphere and the coldest points in the center of the spot. Because of this, filling factors are likely temperature dependent, and the approximation of a two-temperature surface results in single-temperature spot filling factors that act as an average across those temperature gradients. The presence of hot spots (e.g, plages) on the stellar surface would result in an artificial bias in the measured starspot filling factors, as their spectroscopic signatures are not discernible from a hotter stellar surface.

Recent work by \citet{Flores2022} suggests that infrared-derived masses provide a better approximation of the true stellar mass than optically derived masses. While we do not posses dynamical masses for our sample stars to provide a direct unbiased comparison, an analysis of the photospheric and spot temperatures from Table~\ref{tab:spotdata} and the unspotted tracks shown in Figure~\ref{fig:hrcomparison} shows that the spot temperatures provide mass values that are in closer agreement with our spot-corrected masses, while photospheric temperatures systematically overestimate the spot-corrected masses. Given that spots dominate more than half of the stellar surface for all but two of the stars, and that the spot temperatures better reproduce the best-fit stellar temperature, the question arises as to whether we are interpreting cooler stars with substantially hotter spots as opposed to hotter stars with extensive starspots on their surface. Our simplistic models cannot discern between either possibility, as their spectroscopic signature is identical. 

\subsection{Could our Results be Explained by Unaccounted Binarity?} \label{sec:binary}
One of the limitations of the two-temperature spectral models is that they cannot distinguish between the combined emission from a hot photosphere and cool starspots, and that coming from an unresolved binary system with hotter and cooler components. However, we can demonstrate that the latter scenario is not physically plausible. Following the methodology of Section~\ref{sec:HR}, we calculated fictitious radii and luminosities for all the sources under the assumption that the spot and photospheric components originate from different stars, as opposed to a single spotted star. The resulting values are shown in Table~\ref{tab:binary}, with the values presented corresponding to the median across all epochs. $R_{phot}^{\star}$ and $L_{phot}^{\star}$ correspond to the fictitious stellar radii and luminosity that the photospheric component in a binary system would need to have to reproduce the observed stellar spectrum, assuming that the component has a temperature equal to the photospheric temperature. Similarly, $R_{spot}^{\star}$ and $L_{spot}^{\star}$ are the fictitious components corresponding the spot.

\begin{deluxetable}{ccccc}[h!]
\tablecaption{Radii and Luminosities for Binary Models}\label{tab:binary}
\tablehead{\colhead{Star}& \colhead{$R_{phot}^{\star}$} & \colhead{$R_{spot}^{\star}$}  & \colhead{$Log\ (L_{phot}^{\star}/L_{\odot})$} & \colhead{$Log\ (L_{spot}^{\star}/L_{\odot})$}\\
\colhead{} & \colhead{$[R_{\odot}$]} & \colhead{[$R_{\odot}$]}}
\decimals
\small
\setlength{\tabcolsep}{2pt}
\startdata
DI Tau & $0.96$ & $1.46$ & $-0.34$ & $-0.61$\\
IW Tau & $1.06$ & $1.62$ & $-0.29$ & $-0.58$\\
JH 108 & $0.80$ & $1.24$ & $-0.58$ & $-0.77$\\
LkCa 4 & $0.88$ & $1.69$ & $-0.44$ & $-0.60$\\
LkCa 7 & $1.09$ & $1.28$ & $-0.35$ & $-0.85$\\
LkCa 14 & $1.21$ & $0.87$ & $-0.24$ & $-1.36$\\
V819 Tau & $1.81$ & $1.51$ & $-0.06$ & $-0.78$\\
V826 Tau & $1.46$ & $1.36$ & $-0.09$ & $-0.83$\\
V830 Tau & $0.86$ & $1.22$ & $-0.42$ & $-0.76$\\
V1075 Tau & $1.13$ & $1.02$ & $-0.27$ & $-1.09$\\
\enddata
\end{deluxetable}

Given the large filling factors found for stars in our sample, it can be the case over some portions of the near-infrared spectrum that the majority of the stellar flux is coming from the spots and not the photopshere of the star. Therefore, we find radii for the binary components corresponding to the spot emission that are often larger than those corresponding to the photospheric component, as their lower temperature requires a larger emitting surface. For other stars, we find more reasonable results where the hotter larger companion provides more of the net stellar flux. Overall however these radii and luminosities are systematically lower than those reported in Table~\ref{tab:spotdata} which are typical for stars in Taurus.
Furthermore, there is no evidence for close, unresolved binaries in our sample with the exception of three unresolved binaries: IW~Tau, V826~Tau, and DI~Tau. In light of this fact and given the unrealistic stellar parameters that would be required for such binary systems to generate the observed spectra, we consider the possibility that our spotted spectral models could be replaced by unresolved binaries is unlikely. Perhaps more troubling is the fact that if the stars were indeed binaries with the parameters calculated here, age differences from both spotted and unspotted tracks (see Figure~\ref{fig:hrcomparison}) between the primary and secondary components would exceed an order of magnitude, suggesting unrealistic age spreads for the Taurus-Auriga star forming region. With this in mind, we move forward under the assumption that unresolved binaries are not responsible for the spot signals we find here.

\section{Summary and Conclusion}\label{sec:conclusion}
Using a multi-epoch dataset of NIR medium-resolution SpeX spectra, we derive a new set of spot-corrected ages and masses for ten WTTS in the Taurus-Auriga star forming region.
We construct two-temperature composite models of spotted stars by combining BTSettl-CIFIST theoretical atmospheres to represent the photosphere and spots while simultaneously accounting for emission from a warm circumstellar disk using a blackbody at 1500~K. Through a Markov-Chain Monte-Carlo algorithm, we constrain best-fit photospheric and spot temperatures, as well as the spot and disk filling factors for every epoch of spectral data. This methodology allows us to simultaneously reconstruct the entire 0.75-2.40~\micron\ NIR spectra for all stars in our sample. Using the constraints for $T_{phot}$, $T_{spot}$, and $f_{spot}$, we calculate effective temperatures that account for the effect of spots. We extend our models over the spectral range between 300~nm and 75~\micron\ and use Gaia distances to calculate stellar luminosities from which we determine the stellar radii. The median ratio of the spot temperature to the photospheric temperature we have found for this sample of stars in Taurus-Auriga is 0.69 with a standard deviation of 0.03. This ratio is systematically lower than that found for more evolved RS~CVn stars \citep{Strassmeier2009} of $\approx0.8${, but is consistent with previous results for cool dwarfs and giants between 3000~K and 4000~K in \citep{Herbst2020}.}

Placing the stars in an HR~diagram alongside an interpolated set of isochrones and evolutionary tracks from evolutionary models for spotted stars \citep{Somers2020}, we infer new spot-corrected ages and masses. We repeated this step for the best-fit single-temperature unspotted models to determine non spot-corrected ages and masses, and compare the results with spot-corrected ages and masses. We find that correcting for the effects of starspots results in stars appearing $34\%$ and $88\%$ more massive with respect to optical and Red-NIR masses, respectively, with median mass differences on the order of $\Delta M = 0.24\ M_{\odot}$ and $\Delta M = 0.44\ M_{\odot}$. In age, we find stars to shift towards older values by $136\%$ and $195\%$ older for optical and Red-NIR ages, respectively, with median age shifts of $\Delta Age = 1.96$~Myr and $\Delta Age = 2.25$~Myr. These differences result in the stellar population of WTTS in Taurus-Auriga shifting from ages between 0.37 and 2.66 Myr to encompass a range between 0.70 and 6.37 Myr. Cumulatively, these results highlight the substantial effect that starspots can have in age and mass determinations for the youngest stars. 

Owing to the large filling factors we find across our sample, starspots can have a substantial contribution to the total stellar luminosity at NIR wavelengths. For this reason, starpots arise as a compelling explanation for the color anomalies seen in young stars, where young stars are systematically redder than their main sequence counterparts \citep[e.g.,][]{Grankin2008, Gullbring1998}. Spectral type mismatches, likewise, can be explained through the overwhelming contribution of spot flux to the stellar flux in select spectral windows. At the same time, the correlation we recover between starspot filling factors and photometric variability provides strong evidence that starspots are to blame for the characteristic variability found in these systems. This is supported by the ability of spotted models to predict the observed V-Band variability amplitudes (as was found by \citet{myself}). As shown here, starspots lower the temperatures we measure for young stars, which in turn lead to large differences in the masses derived using distinct spectral windows, with both optical and Red-NIR regimes underestimating stellar masses. Cumulatively, the results found herein highlight the substantial impact that can result from a failure to account for starspots. Therefore, it is crucial that the effect of starspots be considered when performing analyses of young stellar systems, in particular when determining fundamental stellar parameters such as effective temperature and luminosity. This problem is exacerbated when using spectra ranging from red-optical to near-infrared where spot contamination is at a maximum.

The process outlined above provides a simple method for determining stellar parameters of spotted stars. Given the large shifts in ages and masses introduced by spots, future efforts to constrain the effects of starspots on the evolution of young stars will benefit from larger surveys of star-forming regions of different ages. Comparisons between masses derived in this way and dynamically-derived masses will provide valuable tests for the validity of the spotted star evolutionary models, and help constrain the spotted nature of these stars.

The authors wish to thank Adolfo Carvalho for insightful mentorship and discussions regarding the implementation of several crucial steps in our analysis and Python algorithms. We also thank the referee for his thorough review and thoughtful suggestions. We thank the staff at NASA's IRTF for their support in conducting the observations that made this work possible. We acknowledge with thanks the variable star observations from the AAVSO International Database contributed by observers worldwide and used in this research. FPP and MM recognize generous travel support through the Volgenau-Wiley Endowed Research Fund at Colgate University. The authors wish to recognize as well the cultural significance the summit of Maunakea holds within the indigenous Hawaiian community. We have been tremendously fortunate for having had the opportunity to conduct observations from this mountain.



\pagebreak
\section{Software and third party data repository citations} \label{sec:cite}
\facilities{IRTF (SpeX)}

\software{astropy \citep{astropy2013}, Spextool \citep{cushing2004}, matplotlib \citep{Hunter2007}, Scipy 
\citep{jones2014}, Numpy \citep{VanderWalt2011}, lmfit \citep{Newville2014}, emcee \citep{Foreman-Mackey2013}, SpectRes \citep{2017arXiv170505165C}, Coronagraph \citep{Coronagraph1, Coronagraph2}}

\restartappendixnumbering
\appendix
\section{Two-temperature Fitting Parameters for Every Epoch.}

\startlongtable
\begin{deluxetable}{ccccccccc}
\tablecaption{Two-Temperature Fitting Results for all Epochs.}\label{tab:appendixtable}
\tablehead{\colhead{Star} & \colhead{MJD} & \colhead{$T_{phot}$} & \colhead{$T_{spot}$}  & \colhead{$f_{spot}$} & \colhead{$f_{disk}$} & \colhead{$T_{eff}$} & \colhead{$R/R_{\odot}$} & \colhead{$Log\ (L/L_{\odot})$}\\
\colhead{} & \colhead{} & \colhead{[K]} & \colhead{[K]} & \colhead{} & \colhead{} & \colhead{[K]}}
\decimals
\startdata
DI Tau & 58489.407 & 4921 & 3399 & 0.81 & 10.31 & 3852 & 2.02 & -0.09 \\
DI Tau & 58490.377 & 4769 & 3358 & 0.77 & 10.49 & 3844 & 1.98 & -0.11 \\
DI Tau & 58491.405 & 5054 & 3440 & 0.83 & 13.95 & 3881 & 1.95 & -0.11 \\
DI Tau & 58493.431 & 4233 & 3160 & 0.56 & 16.88 & 3745 & 2.31 & -0.03 \\
IW Tau & 59864.476 & 4609 & 3296 & 0.68 & 0.04 & 3869 & 1.99 & -0.10 \\
IW Tau & 59864.539 & 4465 & 3393 & 0.35 & 0.02 & 4180 & 1.83 & -0.03 \\
IW Tau & 59864.599 & 4612 & 3297 & 0.77 & 5.98 & 3735 & 2.29 & -0.04 \\
IW Tau & 59864.629 & 4690 & 3253 & 0.78 & 9.35 & 3732 & 2.33 & -0.02 \\
IW Tau & 59867.421 & 4768 & 3257 & 0.81 & 5.3 & 3704 & 1.92 & -0.20 \\
IW Tau & 59867.607 & 4789 & 3270 & 0.82 & 1.95 & 3711 & 2.05 & -0.15 \\
IW Tau & 59868.377 & 4963 & 3291 & 0.87 & 1.27 & 3678 & 2.16 & -0.11 \\
IW Tau & 59868.513 & 4751 & 3262 & 0.80 & 0.03 & 3732 & 2.41 & 0.01 \\
IW Tau & 59870.400 & 4802 & 3234 & 0.80 & 0.84 & 3729 & 2.00 & -0.16 \\
IW Tau & 59870.477 & 4752 & 3223 & 0.79 & 10.13 & 3719 & 2.11 & -0.12 \\
IW Tau & 59870.591 & 4967 & 3228 & 0.83 & 20.31 & 3730 & 2.52 & 0.05 \\
IW Tau & 59871.421 & 4746 & 3188 & 0.78 & 2.87 & 3715 & 2.11 & -0.12 \\
IW Tau & 59871.535 & 4701 & 3204 & 0.78 & 5.97 & 3705 & 2.03 & -0.15 \\
IW Tau & 59871.645 & 4905 & 3265 & 0.82 & 9.67 & 3740 & 2.11 & -0.11 \\
JH 108 & 59864.454 & 4446 & 3302 & 0.78 & 0.41 & 3663 & 1.59 & -0.39 \\
JH 108 & 59864.511 & 4569 & 3345 & 0.74 & 0.2 & 3784 & 1.59 & -0.33 \\
JH 108 & 59864.577 & 4404 & 3357 & 0.71 & 0.39 & 3758 & 1.62 & -0.32 \\
JH 108 & 59867.556 & 4691 & 3319 & 0.77 & 0.34 & 3776 & 1.61 & -0.32 \\
JH 108 & 59868.442 & 4998 & 3414 & 0.89 & 11.44 & 3709 & 1.56 & -0.38 \\
JH 108 & 59868.604 & 5434 & 3427 & 0.92 & 12.7 & 3745 & 1.34 & -0.50 \\
JH 108 & 59870.443 & 4952 & 3378 & 0.87 & 19.01 & 3729 & 1.43 & -0.45 \\
JH 108 & 59870.562 & 4847 & 3377 & 0.86 & 19.8 & 3714 & 1.83 & -0.24 \\
LkCa 4 & 58489.214 & 4789 & 3157 & 0.85 & 7.88 & 3584 & 1.99 & -0.23 \\
LkCa 4 & 58489.309 & 4857 & 3171 & 0.86 & 5.18 & 3588 & 2.10 & -0.18 \\
LkCa 4 & 58489.396 & 4847 & 3132 & 0.86 & 5.58 & 3546 & 2.06 & -0.22 \\
LkCa 4 & 58490.208 & 4893 & 3049 & 0.90 & 7.65 & 3403 & 1.85 & -0.38 \\
LkCa 4 & 58490.275 & 4664 & 2996 & 0.87 & 7.0 & 3381 & 2.26 & -0.22 \\
LkCa 4 & 58491.239 & 4875 & 3179 & 0.87 & 9.9 & 3568 & 1.92 & -0.27 \\
LkCa 4 & 58491.305 & 4636 & 3129 & 0.84 & 4.5 & 3534 & 2.12 & -0.20 \\
LkCa 4 & 58491.421 & 4827 & 3172 & 0.85 & 6.38 & 3586 & 2.09 & -0.18 \\
LkCa 4 & 58492.222 & 5323 & 3301 & 0.90 & 10.37 & 3715 & 2.05 & -0.14 \\
LkCa 4 & 58492.297 & 4716 & 3140 & 0.80 & 5.4 & 3639 & 2.04 & -0.18 \\
LkCa 4 & 58493.205 & 5063 & 3081 & 0.91 & 5.87 & 3458 & 3.07 & 0.08 \\
LkCa 4 & 58493.293 & 4560 & 3041 & 0.86 & 6.37 & 3401 & 2.05 & -0.29 \\
LkCa 7 & 58489.299 & 4524 & 3133 & 0.67 & 5.33 & 3781 & 1.79 & -0.23 \\
LkCa 14 & 58490.359 & 4631 & 2855 & 0.40 & 8.88 & 4167 & 1.65 & -0.13 \\
LkCa 14 & 58491.387 & 4573 & 2871 & 0.38 & 20.13 & 4151 & 1.39 & -0.28 \\
LkCa 14 & 58492.375 & 4573 & 2833 & 0.42 & 8.28 & 4093 & 1.59 & -0.19 \\
LkCa 14 & 58493.365 & 4559 & 2805 & 0.45 & 6.82 & 4043 & 1.58 & -0.22 \\
V819 Tau & 54045.000 & 4151 & 2999 & 0.46 & 21.19 & 3752 & 0.02 & 2.42 \\
V826 Tau & 58489.232 & 4451 & 2894 & 0.52 & 14.52 & 3869 & 2.05 & -0.07 \\
V826 Tau & 58489.328 & 4439 & 3029 & 0.53 & 8.38 & 3886 & 2.19 & 0.00 \\
V826 Tau & 58489.429 & 4530 & 3168 & 0.51 & 13.45 & 4002 & 2.20 & 0.05 \\
V826 Tau & 58490.224 & 4453 & 2959 & 0.53 & 7.99 & 3881 & 2.31 & 0.04 \\
V826 Tau & 58490.290 & 4416 & 2964 & 0.51 & 17.64 & 3878 & 2.21 & 0.00 \\
V826 Tau & 58490.392 & 4481 & 3133 & 0.52 & 7.29 & 3945 & 2.13 & -0.01 \\
V826 Tau & 58491.254 & 4497 & 3112 & 0.50 & 12.61 & 3979 & 2.08 & -0.01 \\
V826 Tau & 58491.321 & 5224 & 3377 & 0.78 & 6.49 & 4046 & 1.91 & -0.06 \\
V826 Tau & 58492.238 & 4574 & 3147 & 0.53 & 14.73 & 4004 & 1.99 & -0.04 \\
V826 Tau & 58492.312 & 4546 & 3158 & 0.59 & 8.16 & 3913 & 2.11 & -0.03 \\
V826 Tau & 58492.434 & 4615 & 3243 & 0.61 & 11.07 & 3949 & 2.15 & 0.00 \\
V826 Tau & 58493.221 & 4524 & 3263 & 0.56 & 14.59 & 3968 & 2.12 & 0.00 \\
V826 Tau & 58493.308 & 4448 & 3122 & 0.54 & 14.26 & 3905 & 2.01 & -0.07 \\
V830 Tau & 59864.428 & 4956 & 3333 & 0.80 & 15.33 & 3849 & 1.56 & -0.32 \\
V830 Tau & 59864.491 & 4880 & 3383 & 0.79 & 6.53 & 3867 & 1.69 & -0.24 \\
V830 Tau & 59867.443 & 4891 & 3371 & 0.77 & 11.84 & 3891 & 1.62 & -0.26 \\
V830 Tau & 59867.631 & 4926 & 3444 & 0.79 & 10.4 & 3912 & 1.60 & -0.27 \\
V830 Tau & 59868.402 & 4911 & 3442 & 0.75 & 12.6 & 3987 & 1.66 & -0.20 \\
V830 Tau & 59870.378 & 4825 & 3244 & 0.76 & 6.48 & 3832 & 1.68 & -0.26 \\
V830 Tau & 59870.513 & 4559 & 3161 & 0.61 & 9.97 & 3897 & 1.80 & -0.17 \\
V1075 Tau & 59864.482 & 4490 & 3081 & 0.55 & 9.97 & 3909 & 1.60 & -0.27 \\
V1075 Tau & 59864.548 & 4674 & 3263& 0.47 & 9.97 & 4179 & 1.53 & -0.19 \\
V1075 Tau & 59867.592 & 4558 & 2890 & 0.53 & 9.97 & 3942 & 1.73 & -0.19 \\
V1075 Tau & 59868.471 & 4851 & 3437 & 0.66 & 9.97 & 4097 & 1.63 & -0.17 \\
V1075 Tau & 59868.583 & 4698 & 3295 & 0.57 & 9.97 & 4083 & 1.43 & -0.29 \\
V1075 Tau & 59870.498 & 4576 & 2931 & 0.51 & 9.97 & 3979 & 1.65 & -0.21 \\
\enddata
\end{deluxetable}
\bibliography{spots_working}{}
\bibliographystyle{aasjournal}


\end{document}